\documentclass[11pt,a4paper]{cernrep}
\usepackage{rep_common}
\usepackage{cancel}
\usepackage{placeins}
\usepackage{booktabs}
\usepackage{slashbox}
\usepackage{amsmath,amssymb,amsthm}
\usepackage{rotating}
\usepackage{multicol}
\usepackage{tabulary}
\usepackage[colorlinks, urlcolor=blue, linkcolor=blue, citecolor=blue, plainpages=false]{hyperref}
\usepackage{float}
\floatstyle{plaintop}
\restylefloat{table}
\usepackage{xargs}
\usepackage{xstring}
\usepackage{lineno}

\usepackage{graphicx}
\usepackage{siunitx}

\usepackage{xspace}
\usepackage{cite}

\usepackage{graphicx,multirow,colortbl}

\newcommand*{\GEANT}{\textsc{Geant}\xspace}
\newcommand{\pbfd}{PbF$_2$}

\begin{document}

\newcounter{ncontacts}
\newcommand{\fcontact}[1]{\StrCount{#1}{,}[\tmp]\setcounter{ncontacts}{\tmp}
  Contact\ifthenelse{\value{ncontacts} > 0}{s}{}: #1}
\newcommand{\feditor}[1]{\StrCount{#1}{,}[\tmp]\setcounter{ncontacts}{\tmp}
  Contact Editor\ifthenelse{\value{ncontacts} > 0}{s}{}: #1}
\newcommandx{\asection}[2][1=NONE]{
  \ifthenelse{\equal{#1}{NONE}}
  {\section{#2}}{\section[#2]{#2\footnote{\feditor{#1}}}}}
\newcommandx{\asubsection}[2][1=NONE]{
  \ifthenelse{\equal{#1}{NONE}}
  {\subsection{#2}}{
    \subsection[#2]{#2\footnote{\fcontact{#1}}}}}
\newcommandx{\asubsubsection}[2][1=NONE]{
  \ifthenelse{\equal{#1}{NONE}}
  {\subsubsection{#2}}{
    \subsubsection[#2]{#2\footnote{\fcontact{#1}}}}}
\newcommandx{\asubsubsubsection}[2][1=NONE]{
  \ifthenelse{\equal{#1}{NONE}}
  {\subsubsubsection{#2}}{
    \subsubsubsection[#2]{#2\footnote{\fcontact{#1}}}}}

\newcommand\snowmass{\begin{center}\rule[-0.2in]{\hsize}{0.01in}\\\rule{\hsize}{0.01in}\\
    \vskip 0.1in Submitted to the  Proceedings of the US Community Study\\
    on the Future of Particle Physics (Snowmass 2021)\\
    \rule{\hsize}{0.01in}\\\rule[+0.2in]{\hsize}{0.01in} \end{center}}

\title{{\normalfont\bfseries\boldmath\huge
      \begin{center}
        Promising Technologies and R\&D Directions for the Future Muon Collider Detectors
      \end{center}
      \vspace*{-40pt}
    }
    {\textnormal{\normalsize \snowmass}}
    {\textnormal{\normalsize
        This is one of the six reports submitted to Snowmass by the International Muon Collider Collaboration. \\
        The Indico subscription page:\\
        \centerline{
          \href{https://indico.cern.ch/event/1130036/}{https://indico.cern.ch/event/1130036/}}
        contains the link to the reports and gives the possibility to subscribe to the papers. \\
        The policy for signatures is that, for each individual report, you can subscribe as "Author" or as "Signatory", defined as follows:
        \begin{itemize}
          \item
                ``Author'' indicates that you did contribute to the results documented in the report in any form, including e.g. by participating to the discussions of the community meeting, sending comments on the drafts, etc, or that you plan to contribute to the future work. The ``Authors'' will appear as such on arXiv.
          \item
                ``Signatory'' means that you express support to the Collaboration effort and endorse the Collaboration plans. The ``Signatories'' list will be reported in the text only.
        \end{itemize}
        The white papers will appear on arXiv on 15 March 2022 (Snowmass deadline).\\ Additional contributors can subscribe until March 30th and they will be added to the revised version.
      }}
}

\newcounter{instituteref}
\newcommand{\iinstitute}[2]{\refstepcounter{instituteref}\label{#1}$^{\ref{#1}}$\href{http://inspirehep.net/record/#1}{#2}}
\newcommand{\iauthor}[3]{\href{http://inspirehep.net/record/#1}{#2}$^{#3}$}
\author{Editors: \\ 
\iauthor{1028687}{{S.~Jindariani}}{\ref{902796}}, 
\iauthor{1074063}{F.~Meloni}{\ref{902770}}, 
\iauthor{994095}{{N.~Pastrone}}{\ref{902889}} 
\\ \vspace*{4mm}Authors: \\ 
\iauthor{1757334}{C.~Aim{\`e}}{\ref{943385},\ref{902885}}, 
\iauthor{1073143}{N.~Bartosik}{\ref{902889}}, 
\iauthor{1037853}{{E.~Barzi}}{\ref{902796},\ref{903092}}, 
\iauthor{1029828}{A.~Bertolin}{\ref{902884}}, 
\iauthor{1015478}{{A.~Braghieri}}{\ref{902885}}, 
\iauthor{1894439}{L.~Buonincontri}{\ref{902884},\ref{903113}}, 
\iauthor{1707397}{{S.~Calzaferri}}{\ref{902885}}, 
\iauthor{1057458}{M.~Casarsa}{\ref{902888}}, 
\iauthor{1014281}{M.G.~Catanesi}{\ref{902877}}, 
\iauthor{1021757}{A.~Cerri}{\ref{1241166}}, 
\iauthor{1037833}{G.~Chachamis}{\ref{1294609}}, 
\iauthor{1013275}{A.~Colaleo}{\ref{902660},\ref{902877}}, 
\iauthor{1937290}{{C.~Curatolo}}{\ref{902882}}, 
\iauthor{2049478}{{G.~Da~Molin}}{\ref{909099}}, 
\iauthor{1012143}{J.~Delahaye}{\ref{902725}}, 
\iauthor{1031269}{B.~Di~Micco}{\ref{906528},\ref{907692}}, 
\iauthor{1011508}{T.~Dorigo}{\ref{902884}}, 
\iauthor{1404358}{F.~Errico}{\ref{902660},\ref{902877}}, 
\iauthor{1719039}{D.~Fiorina}{\ref{902885}}, 
\iauthor{1262268}{A.~Gianelle}{\ref{902884}}, 
\iauthor{1971617}{{C.~Giraldin}}{\ref{903113}}, 
\iauthor{2044726}{J.~Hauptman}{\ref{902893}}, 
\iauthor{1067690}{T.R.~Holmes}{\ref{1623978}}, 
\iauthor{1252769}{K.~Krizka}{\ref{902953}}, 
\iauthor{1071846}{L.~Lee}{\ref{1623978}}, 
\iauthor{999862}{K.R.~Long}{\ref{902868},\ref{903174}}, 
\iauthor{999654}{D.~Lucchesi}{\ref{903113},\ref{902884}}, 
\iauthor{997065}{N.~Mokhov}{\ref{902796}}, 
\iauthor{2049482}{A.~Montella}{\ref{902888}}, 
\iauthor{}{F.~Nardi}{\ref{903113},\ref{902884}}, 
\iauthor{995827}{N.~Neufeld}{\ref{902725}}, 
\iauthor{995826}{{D.~Neuffer}}{\ref{902796}}, 
\iauthor{1048820}{S.~Pagan~Griso}{\ref{902953}}, 
\iauthor{1772198}{A.~Pellecchia}{\ref{902660}}, 
\iauthor{1050691}{K.~Potamianos}{\ref{903112}}, 
\iauthor{992463}{E.~Radicioni}{\ref{902877}}, 
\iauthor{1217056}{R.~Radogna}{\ref{902660},\ref{902877}}, 
\iauthor{1020819}{C.~Riccardi}{\ref{943385},\ref{902885}}, 
\iauthor{1028713}{L.~Ristori}{\ref{902796}}, 
\iauthor{991185}{L.~Rossi}{\ref{903009},\ref{907142}}, 
\iauthor{990505}{P.~Salvini}{\ref{902885}}, 
\iauthor{989645}{D.~Schulte}{\ref{902725}}, 
\iauthor{1342183}{L.~Sestini}{\ref{902884}}, 
\iauthor{988978}{V.~Shiltsev}{\ref{902796}}, 
\iauthor{1622677}{F.~M.~Simone}{\ref{902660},\ref{902877}}, 
\iauthor{}{A.~Stamerra}{\ref{902660},\ref{902877}}, 
\iauthor{1071725}{X.~Sun}{\ref{1210798}}, 
\iauthor{1268914}{J.~Tang}{\ref{1283410},\ref{903123}}, 
\iauthor{1878399}{E.~A.~Thompson}{\ref{902770}}, 
\iauthor{1265350}{I.~Vai}{\ref{902885}}, 
\iauthor{1643523}{N.~Valle}{\ref{943385},\ref{902885}}, 
\iauthor{1071756}{R.~Venditti}{\ref{902660},\ref{902877}}, 
\iauthor{1063935}{P.~Verwilligen}{\ref{902877}}, 
\iauthor{984555}{P.~~Vitulo}{\ref{943385},\ref{902885}}, 
\iauthor{1073818}{{H.~Weber}}{\ref{902858}}, 
\iauthor{1019845}{K.~Yonehara}{\ref{902796}}, 
\iauthor{1971310}{A.~Zaza}{\ref{902660},\ref{902877}}, 
\iauthor{1863481}{D.~Zuliani}{\ref{903113},\ref{902884}} 
\\ \vspace*{4mm} Signatories: \\ 
\iauthor{1018902}{K.~Agashe}{\ref{902990}}, 
\iauthor{1018633}{B.C.~Allanach}{\ref{907623}}, 
\iauthor{1491320}{P.~Asadi}{\ref{1237813}}, 
\iauthor{1067349}{{M.A.~Mahmoud}.}{\ref{912409}}, 
\iauthor{1041900}{A.~Azatov}{\ref{4416},\ref{902888}}, 
\iauthor{2031609}{F.~Batsch}{\ref{902725}}, 
\iauthor{1049763}{A.~Bersani}{\ref{902881}}, 
\iauthor{1037456}{M.E.~Biagini}{\ref{902807}}, 
\iauthor{1020223}{K.M.~Black}{\ref{903349}}, 
\iauthor{}{{S.A.~Bogacz}}{\ref{99999}}, 
\iauthor{1015872}{M.~Bonesini}{\ref{902882},\ref{907960}}, 
\iauthor{1015214}{A.~Bross}{\ref{902796}}, 
\iauthor{1077579}{D.~Buttazzo}{\ref{902886}}, 
\iauthor{1670912}{B.~Caiffi}{\ref{902881}}, 
\iauthor{1014742}{G.~Calderini}{\ref{926589},\ref{903119}}, 
\iauthor{1275234}{R.~Capdevilla}{\ref{908474},\ref{903282}}, 
\iauthor{}{L.~Castelli}{\ref{903113}}, 
\iauthor{1793525}{C.~Cesarotti}{\ref{902835}}, 
\iauthor{1014143}{Z.~Chacko}{\ref{902990}}, 
\iauthor{2023221}{A.~Chanc\'e}{\ref{912490}}, 
\iauthor{2037614}{S.~Chen}{\ref{1471035}}, 
\iauthor{1069708}{Y.-T.~Chien}{\ref{1275736}}, 
\iauthor{1862239}{M.~Costa}{\ref{903128},\ref{902886}}, 
\iauthor{1046385}{N.~Craig}{\ref{903307}}, 
\iauthor{1024481}{D.~Curtin}{\ref{903282}}, 
\iauthor{1067364}{R.~T.~D'Agnolo}{\ref{1087875}}, 
\iauthor{}{{M.~Dam}}{\ref{902882}}, 
\iauthor{1076225}{H.~Damerau}{\ref{902725}}, 
\iauthor{1012395}{S.~Dasu}{\ref{903349}}, 
\iauthor{1021004}{J.~de~Blas}{\ref{903836}}, 
\iauthor{1651018}{{E.~De~Matteis}}{\ref{907142}}, 
\iauthor{1012237}{A.~Deandrea}{\ref{1743848}}, 
\iauthor{1019723}{A.~Delgado}{\ref{903085}}, 
\iauthor{1012025}{H.~Denizli}{\ref{908452}}, 
\iauthor{1011983}{R.~Dermisek}{\ref{902874}}, 
\iauthor{1395010}{K.~F.~Di~Petrillo}{\ref{902796}}, 
\iauthor{1246709}{J.~Dickinson}{\ref{902796}}, 
\iauthor{1064320}{P.~Everaerts}{\ref{903349}}, 
\iauthor{1010523}{L.~Everett}{\ref{903349}}, 
\iauthor{1069878}{S.~Farinon}{\ref{902881}}, 
\iauthor{1010065}{G.~Ferretti}{\ref{902825}}, 
\iauthor{1009979}{F.~Filthaut}{\ref{903075}}, 
\iauthor{1894571}{{M.~Forslund}}{\ref{910429}}, 
\iauthor{1052115}{{R.~Franceschini}}{\ref{906528},\ref{907692}}, 
\iauthor{1009120}{E.~Gabrielli}{\ref{903287},\ref{902888}}, 
\iauthor{1946817}{F.~Garosi}{\ref{904416}}, 
\iauthor{1894454}{L.~Giambastiani}{\ref{1513358},\ref{902884}}, 
\iauthor{1706734}{A.~Glioti}{\ref{1471035}}, 
\iauthor{1059457}{L.~Gray}{\ref{902796}}, 
\iauthor{1198373}{A.~Greljo}{\ref{902668}}, 
\iauthor{1007486}{C.~Grojean}{\ref{902770},\ref{902858}}, 
\iauthor{1274618}{J.~Gu}{\ref{903628}}, 
\iauthor{1259916}{C.~Han}{\ref{903702}}, 
\iauthor{1006825}{T.~Han}{\ref{903130}}, 
\iauthor{1383268}{B.~Henning}{\ref{1471035}}, 
\iauthor{1912097}{{K.~Hermanek}}{\ref{902874}}, 
\iauthor{1515880}{S.~Homiller}{\ref{902835}}, 
\iauthor{1475406}{S.~Jana}{\ref{902841}}, 
\iauthor{2049476}{H.~Jia}{\ref{903349}}, 
\iauthor{1051663}{Y.~Kahn}{\ref{902867}}, 
\iauthor{1003695}{D.~M.~Kaplan}{\ref{1273767}}, 
\iauthor{1002991}{W.~Kilian}{\ref{903203}}, 
\iauthor{1020007}{K.~Kong}{\ref{902912}}, 
\iauthor{1019544}{P.~Koppenburg}{\ref{903832}}, 
\iauthor{1345391}{G.K.~Krintiras}{\ref{1273495}}, 
\iauthor{1064657}{W.~Li}{\ref{903156}}, 
\iauthor{1074984}{Q.~Li}{\ref{903603}}, 
\iauthor{1256188}{Z.~Liu}{\ref{903010}}, 
\iauthor{1700371}{{Q.~Lu}}{\ref{902835}}, 
\iauthor{1514492}{Y.~Ma}{\ref{903130}}, 
\iauthor{999053}{F.~Maltoni}{\ref{910783},\ref{902674}}, 
\iauthor{1668860}{L.~Mantani}{\ref{907623}}, 
\iauthor{1670119}{{S.~Mariotto}}{\ref{903009},\ref{907142}}, 
\iauthor{1078065}{D.~Marzocca}{\ref{902888}}, 
\iauthor{1971307}{P.~Mastrapasqua}{\ref{910783}}, 
\iauthor{998430}{K.~Matchev}{\ref{902804}}, 
\iauthor{1054925}{A.~Mazzacane}{\ref{902796}}, 
\iauthor{1751811}{N.~McGinnis}{\ref{903290}}, 
\iauthor{1025277}{P.~Meade}{\ref{910429}}, 
\iauthor{997877}{{B.~Mele}}{\ref{902887}}, 
\iauthor{1022138}{P.~Merkel}{\ref{902796}}, 
\iauthor{1461119}{C.~Merlassino}{\ref{903112}}, 
\iauthor{1066275}{R.~K.~Mishra}{\ref{902835}}, 
\iauthor{1070072}{A.~Mohammadi}{\ref{903349}}, 
\iauthor{}{R.~Mohapatra}{\ref{99999}}, 
\iauthor{996989}{P.~Montagna}{\ref{943385},\ref{902885}}, 
\iauthor{1064125}{R.~Musenich}{\ref{902881}}, 
\iauthor{995460}{{Y.~Nomura}}{\ref{903299}}, 
\iauthor{1070110}{I.~Ojalvo}{\ref{16750}}, 
\iauthor{1077958}{P.~Panci}{\ref{903129},\ref{902886}}, 
\iauthor{1023838}{P.~Paradisi}{\ref{1513358},\ref{902884}}, 
\iauthor{1067962}{{A.~Perloff}}{\ref{902748}}, 
\iauthor{993440}{F.~Piccinini}{\ref{902885}}, 
\iauthor{1021028}{M.~Pierini}{\ref{902725}}, 
\iauthor{1651162}{{M.~Prioli}}{\ref{907142}}, 
\iauthor{1024769}{M.~Procura}{\ref{903326}}, 
\iauthor{992212}{T.~Raubenheimer}{\ref{903206}}, 
\iauthor{1214912}{D.~Redigolo}{\ref{1214912}}, 
\iauthor{992031}{L.~Reina}{\ref{902803}}, 
\iauthor{1021811}{J.~Reuter}{\ref{902770}}, 
\iauthor{1056642}{F.~Riva}{\ref{902813}}, 
\iauthor{1040385}{T.~Robens}{\ref{902678}}, 
\iauthor{1072232}{F.~Sala}{\ref{908583}}, 
\iauthor{1885424}{J.~Salko}{\ref{902668}}, 
\iauthor{1077871}{E.~Salvioni}{\ref{1513358},\ref{902884}}, 
\iauthor{990367}{J.~Santiago}{\ref{909079},\ref{903836}}, 
\iauthor{}{I.~Sarra}{\ref{99999}}, 
\iauthor{989950}{J.~Schieck}{\ref{903324},\ref{904536}}, 
\iauthor{1039590}{M.~Selvaggi}{\ref{902725}}, 
\iauthor{1019799}{A.~Senol}{\ref{908452}}, 
\iauthor{1071696}{V.~Sharma}{\ref{903349}}, 
\iauthor{1066476}{M.~Sorbi}{\ref{903009},\ref{907142}}, 
\iauthor{1319078}{G.~Stark}{\ref{1218068}}, 
\iauthor{1057643}{M.~Statera}{\ref{907142}}, 
\iauthor{1071880}{J.~Stupak}{\ref{1273509}}, 
\iauthor{987285}{S.~Su}{\ref{902647}}, 
\iauthor{987128}{{R.~Sundrum}}{\ref{912511}}, 
\iauthor{1078570}{M.~Swiatlowski}{\ref{903290}}, 
\iauthor{1064514}{R.~Torre}{\ref{902881}}, 
\iauthor{985810}{L.~~Tortora}{\ref{907692},\ref{902725}}, 
\iauthor{1778841}{S.~Trifinopoulos}{\ref{902888}}, 
\iauthor{1613622}{M.~Valente}{\ref{903290}}, 
\iauthor{2025179}{R.~U.~Valente}{\ref{907142}}, 
\iauthor{1863232}{{L.~Vittorio}}{\ref{903128},\ref{902886}}, 
\iauthor{1077733}{E.~~Vryonidou}{\ref{902984}}, 
\iauthor{1054127}{C.~Vuosalo}{\ref{903349}}, 
\iauthor{1049718}{M.~Wendt}{\ref{902725}}, 
\iauthor{1511975}{{Y.~Wu}}{\ref{903094}}, 
\iauthor{1037622}{{A.~Wulzer}}{\ref{909099}}, 
\iauthor{1618109}{K.~Xie}{\ref{903130}}, 
\iauthor{1024759}{{H.-B.~Yu}}{\ref{903304}}, 
\iauthor{1066114}{Y.~J.~Zheng}{\ref{902912}}, 
\iauthor{1037623}{J.~Zurita}{\ref{907907}} 
\vspace*{1cm}} \institute{\small 
\iinstitute{902796}{{Fermi National Accelerator Laboratory, Batavia Illinois 60510-5011, USA}, United States}; 
\iinstitute{902770}{Deutsches Elektronen-Synchrotron DESY, Notkestr. 85, 22607 Hamburg, Germany}; 
\iinstitute{902889}{{Istituto Nazionale di Fisica Nucleare, Sezione di Torino via P. Giuria, 1, 10125 Torino,  Italy}}; 
\iinstitute{943385}{Universit{\`a} di Pavia, Italy}; 
\iinstitute{902884}{INFN Sezione di Padova, Padova, Italy}; 
\iinstitute{902885}{{INFN, Sezione di Pavia, via A. Bassi 6, I-27100  Pavia, Italy}}; 
\iinstitute{902888}{INFN Sezione di Trieste, Trieste, Italy}; 
\iinstitute{902877}{INFN , Bari, Italy}; 
\iinstitute{1241166}{MPS School, University of Sussex, Sussex House, BN19QH Brighton, United Kingdom}; 
\iinstitute{1294609}{Laborat{\' o}rio de Instrumenta\c{c}{\~ a}o e F{\' \i}sica Experimental de Part{\' \i}culas (LIP),  Av. Prof. Gama Pinto, 2, P-1649-003 Lisboa, Portugal}; 
\iinstitute{902660}{{Department of Physics, Universit{\`a} degli Studi di Bari, Italy}}; 
\iinstitute{902882}{{INFN, Sezione di Milano, via Celoria 16, 20133 Milan, Italy}}; 
\iinstitute{909099}{{Dipartimento di Fisica e Astronomia, Universit\'a di Padova}, Italy}; 
\iinstitute{902725}{CERN, Switzerland}; 
\iinstitute{906528}{Università degli Studi di Roma Tre, Italy}; 
\iinstitute{903113}{{Dipartimento di Fisica, Universit{\`a} degli Studi di Padova, via Marzolo 8, Padova, Italy}}; 
\iinstitute{902893}{{Iowa State University, Ames, Iowa,  50011 USA}, United States}; 
\iinstitute{1623978}{{University of Tennessee, Knoxville, TN, USA}, United States}; 
\iinstitute{902953}{{Physics Division, Lawrence Berkeley National Laboratory, Berkeley, CA, USA, United States}}; 
\iinstitute{902868}{Imperial College London, Exhibition Road, London, SW7 2AZ, UK, United Kingdom}; 
\iinstitute{903112}{Department of Physics, Oxford University, Oxford, United Kingdom}; 
\iinstitute{903009}{{Dipartimento di Fisica Aldo Pontremoli, Università degli Studi di Milano, Via Celoria, 16, 20133 Milano, Italy}}; 
\iinstitute{1210798}{State Key Laboratory of Nuclear Physics and Technology, Peking University, Beijing, China}; 
\iinstitute{1283410}{Sun Yat-sen University, guangzhou 510275, China}; 
\iinstitute{902858}{{Institut f{\"u}r Physik, Humboldt-Universit{\"a}t zu Berlin, Newstonstr.\ 15, 12489 Berlin, Germany}}; 
\iinstitute{903092}{{Ohio State University, Columbus, OH 43210}, United States}; 
\iinstitute{907692}{Istituto Nazionale di Fisica Nucleare sezione di Roma Tre, Italy}; 
\iinstitute{903174}{STFC, United Kingdom}; 
\iinstitute{907142}{{Laboratorio Acceleratori e Superconduttività Applicata (LASA), Istituto Nazionale di Fisica Nucleare (INFN), Via Fratelli Cervi 201, 20054 Segrate Milano, Italy}}; 
\iinstitute{903123}{Institute of High-Energy Physics, Beijing, China}; 
\iinstitute{902990}{Maryland Center for Fundamental Physics, University of Maryland, College Park, MD 20742, USA, United States}; 
\iinstitute{907623}{DAMTP, Wilberforce Road, University of Cambridge, Cambridge CB3 0WA, United Kingdom}; 
\iinstitute{1237813}{Center for Theoretical Physics, Massachusetts Institute of Technology,   Cambridge, MA 02139, USA., United States}; 
\iinstitute{912409}{{Center for High Energy Physics (CHEP-F), Fayoum University{\"a}, 63514 El-Fayoum, Egypt}}; 
\iinstitute{4416}{SISSA International School for Advanced Studies, Via Bonomea 265, 34136, Trieste, Italy}; 
\iinstitute{902881}{Istituto Nazionale di Fisica Nucleare - Sezione di Genova - via Dodecaneso 33, 16146 Genova - Italy}; 
\iinstitute{902807}{INFN, Frascati National Laboratory, Italy}; 
\iinstitute{903349}{University of Wisconsin-Madison, United States}; 
\iinstitute{99999}{{Center for Advanced Studies of Accelerators, Jefferson Lab, Newport News, VA 23606, USA}, United States}; 
\iinstitute{902886}{INFN Sezione di Pisa, Largo B.\ Pontecorvo 3, 56127 Pisa, Italy}; 
\iinstitute{926589}{CNRS/IN2P3, Paris, France}; 
\iinstitute{908474}{Perimeter Institute, Canada}; 
\iinstitute{902835}{Department of Physics, Harvard University, Cambridge, MA, 02138, United States}; 
\iinstitute{912490}{IRFU, CEA, Université Paris-Saclay, Gif-sur-Yvette; France}; 
\iinstitute{1471035}{{Theoretical Particle Physics Laboratory (LPTP), Institute of Physics, EPFL, Lausanne, Switzerland}}; 
\iinstitute{1275736}{Physics and Astronomy Department, Georgia State University, Atlanta, GA 30303, U.S.A., United States}; 
\iinstitute{903128}{Scuola Normale Superiore, Piazza dei Cavalieri 7, 56126 Pisa, Italy}; 
\iinstitute{903307}{University of California, Santa Barbara, United States}; 
\iinstitute{903282}{Department of Physics, University of Toronto, Canada}; 
\iinstitute{1087875}{Universit\`e Paris Saclay, CNRS, CEA, Institut de Physique Th\`eorique, 91191, Gif-sur-Yvette, France}; 
\iinstitute{903836}{CAFPE and Departamento de F\'isica Te\'orica y del Cosmos, Universidad de Granada, Granada, Spain}; 
\iinstitute{1743848}{IP2I, Universit\'e Lyon 1, CNRS/IN2P3, UMR5822, F-69622, Villeurbanne, France}; 
\iinstitute{903085}{University of Notre Dame, United States}; 
\iinstitute{908452}{Department of Physics, Bolu Abant Izzet Baysal University, 14280, Bolu, Turkey}; 
\iinstitute{902874}{Physics Department, Indiana University, Bloomington, IN, 47405, USA, United States}; 
\iinstitute{902825}{Chalmers University of Technology 40220 Gothenburg , Sweden}; 
\iinstitute{903075}{Radboud University and Nikhef, Nijmegen, The Netherlands}; 
\iinstitute{910429}{{C. N. Yang Institute for Theoretical Physics, Stony Brook University, Stony Brook, NY 11794, USA}, United States}; 
\iinstitute{903287}{Physics Department, University of Trieste, Strada Costiera 11, 34151 Trieste, Italy}; 
\iinstitute{904416}{SISSA, Italy}; 
\iinstitute{1513358}{Dipartimento di Fisica e Astronomia, Universit\`a degli Studi di Padova, Italy}; 
\iinstitute{902668}{Albert Einstein Center for Fundamental Physics, Institute for Theoretical Physics, University of Bern, CH-3012 Bern, Switzerland}; 
\iinstitute{903628}{Department of Physics, Fudan University, Shanghai 200438, China}; 
\iinstitute{903702}{School of Physics, Sun Yat-Sen University, Guangzhou 510275, China}; 
\iinstitute{903130}{University of Pittsburgh, United States}; 
\iinstitute{902841}{Max-Planck-Institut f{\"u}r Kernphysik, Germany}; 
\iinstitute{902867}{{Department of Physics, University of Illinois at Urbana-Champaign, Urbana, IL 61801, USA}, United States}; 
\iinstitute{1273767}{Illinois Institute of Technology, United States}; 
\iinstitute{903203}{Department of Physics, University of Siegen, 57068 Siegen, Germany}; 
\iinstitute{902912}{Department of Physics and Astronomy, University of Kansas, Lawrence, KS 66045, USA, United States}; 
\iinstitute{903832}{Nikhef National Institute for Subatomic Physics, Amsterdam, The Netherlands}; 
\iinstitute{1273495}{{Department of Physics and Astronomy 1082 Malott, 1251 Wescoe Hall Dr. Lawrence, KS 66045}, United States}; 
\iinstitute{903156}{Rice University, Houston, TX 77005, USA, United States}; 
\iinstitute{903603}{Peking University, Beijing, China}; 
\iinstitute{903010}{{School of Physics and Astronomy, University of Minnesota, Minneapolis, MN 55455, USA}, United States}; 
\iinstitute{910783}{Center for Cosmology, Particle Physics and Phenomenology, Universit\'e catholique de Louvain, B-1348 Louvain-la-Neuve, Belgium}; 
\iinstitute{902804}{Physics Department, University of Florida, Gainesville FL 32611 USA, United States}; 
\iinstitute{903290}{TRIUMF, 4004 Westbrook Mall, Vancouver, BC, Canada V6T 2A3}; 
\iinstitute{902887}{{INFN, Sezione di Roma, c/o Dipartimento di Fisica - Sapienza Universit{\`a} di Roma, Piazzale Aldo Moro, 2, I-00185 Rome, Italy}}; 
\iinstitute{903299}{{Department of Physics, University of California, Berkeley, CA 94720, USA}, United States}; 
\iinstitute{16750}{Princeton University, United States}; 
\iinstitute{903129}{Pisa University, Italy}; 
\iinstitute{902748}{{Department of Physics, University of Colorado, 390 UCB, Boulder, CO 80309, United States}}; 
\iinstitute{903326}{University of Vienna, Faculty of Physics, Boltzmanngasse 5, 1090 Vienna, Austria}; 
\iinstitute{903206}{SLAC National Accelerator Laboratory, United States}; 
\iinstitute{1214912}{INFN Florence, Italy}; 
\iinstitute{902803}{Florida State University, United States}; 
\iinstitute{902813}{D\'epartment de Physique Th\'eorique, Universit\'e de Gen\`eve, 24 quai Ernest-Ansermet, 1211 Gen\`eve 4, Switzerland}; 
\iinstitute{902678}{Rudjer Boskovic Institute, Zagreb, Croatia}; 
\iinstitute{908583}{Laboratoire de Physique Th\'eorique et Hautes \'Energies, Sorbonne Universit\'e, CNRS, Paris, France}; 
\iinstitute{909079}{{CAFPE}, Spain}; 
\iinstitute{903324}{Institut f\"ur Hochenergiephysik der \"Osterreichischen Akademie der Wissenschaften, Nikolsdorfer Gasse 18, A-1050 Wien, Austria}; 
\iinstitute{1218068}{SCIPP, UC Santa Cruz, United States}; 
\iinstitute{1273509}{University of Oklahoma, United States}; 
\iinstitute{902647}{University of Arizona, United States}; 
\iinstitute{912511}{{Maryland Center for Fundamental Physics, University of Maryland, College Park, MD 20742, USA}, United States}; 
\iinstitute{902984}{University of Manchester, Manchester M13 9PL, UK, United Kingdom}; 
\iinstitute{903094}{{Department of Physics, Oklahoma State University, Stillwater, OK, 74078, USA}, United States}; 
\iinstitute{903304}{{Department of Physics and Astronomy, University of California, Riverside, CA 92521, United States}}; 
\iinstitute{907907}{{Instituto de F{\'i}sica Corpuscular, CSIC-Universitat de Val{\'e}ncia, Valencia, Spain}}; 
\iinstitute{907960}{Dipartimento di Fisica Universit\'a Milano Bicocca, Italy}; 
\iinstitute{903119}{LPNHE, Sorbonne Universit\'e, France}; 
\iinstitute{902674}{Dipartimento di Fisica e Astronomia, Università di Bologna,  via Irnerio 46,  I-40126 Bologna  , Italy}; 
\iinstitute{904536}{Atominstitut, Technische Universit\"at Wien, Stadionallee 2,  A-1020 Wien, Austria} 
}

\begin{titlepage}

  \vspace*{-1.8cm}

  \noindent
  \begin{tabular*}{\linewidth}{lc@{\extracolsep{\fill}}r@{\extracolsep{0pt}}}
    \vspace*{-1.2cm}\mbox{\!\!\!\includegraphics[width=.14\textwidth]{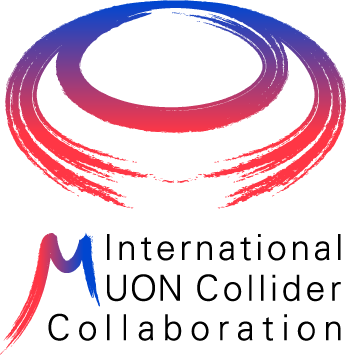}} & &  \\
    & & \today \\  
    & & \href{https://muoncollider.web.cern.ch}{https://muoncollider.web.cern.ch} \\ 
    & & \\
    \hline
  \end{tabular*}

  \vspace*{0.3cm}

  \maketitle

  \begin{abstract}
    Among the post-LHC generation of particle accelerators, the muon collider represents a unique machine with capability to provide very high energy leptonic collisions and to open the path to a vast and mostly unexplored physics programme. However, on the experimental side, such great physics potential is accompanied by unprecedented technological challenges, due to the fact that muons are unstable particles. Their decay products interact with the machine elements and produce an intense flux of background particles that eventually reach the detector and may degrade its performance. In this paper, we present technologies that have a potential to match the challenging specifications of a muon collider detector and outline a path forward for the future R\&D efforts.
  \end{abstract}

\end{titlepage}

\clearpage

\section{Introduction}
\label{sec:intro}

The Higgs boson, discovered at the Large Hadron Collider in 2012 was the last missing piece of the Standard Model. However, the quest for understanding the fundamental laws of nature continues. There is no clear indication of where lie the answers to fundamental questions such as what are the origins of dark matter or the Universe's matter-antimatter imbalance. This calls for an even broader experimental programme aimed at investigating subnuclear physics processes in all the accessible phase space.

While the LHC still has to enter its high-luminosity phase, with several years of data taking to come, work is ongoing to define the long-term strategy for the field and to evaluate the potential of the possible future collider facilities~\cite{sm21-imcc-accelerator-summary,Delahaye:2019omf,Long:2020wfp}.
Most proposals for future colliders rely on either precision $e^{+}e^{-}$ machines or high energy proton--proton collisions to directly probe the energy frontier. A future high-energy muon collider could cover large regions of parameters space not reachable by the LHC, while providing significantly cleaner final states than those produced in hadron collisions.
Even though the idea of a muon collider is not new, as it was first proposed in the late 1960s~\cite{tikhonin1968effects, budker1982accelerators} and studied in detail by the Muon Accelerator Program (MAP)~\cite{Delahaye:2013jla}, it is now receiving renewed interest because of its potential to overcome key limitations of other proposed collider concepts.
Thanks to their mass about 200 times heavier than that of the electron, muons are impervious to synchrotron radiation. As a result, circular muon colliders could reach the multi-TeV regime, with a limited energy footprint. Furthermore, muon collisions are expected to take place with a small energy spread and lead to an improved energy resolution for physics measurements.

The enormous physics potential of colliding muon beams has sparked a wave of studies aimed at defining in detail a possible physics programme~\cite{sm21-imcc-physics-summary,sm21-imcc-physics-3TeV,Buttazzo:2018qqp,Franceschini:2021dxn,Capdevilla:2021fmj,AlAli:2021let,Buttazzo:2020uzc,Han:2020pif,Capdevilla:2020qel,Ruhdorfer:2019utl,Chiesa:2020awd,Costantini:2020stv,Capdevilla:2021rwo,Bartosik:2020xwr,Yin:2020afe,Kalinowski:2020rmb,Liu:2021jyc,Han:2021udl,Bottaro:2021srh,Li:2021lnz,Asadi:2021gah,Sahin:2021xzt,Chen:2021rid,Haghighat:2021djz,Bottaro:2021snn,Sen:2021fha,Han:2021lnp,Bandyopadhyay:2021pld,Dermisek:2021mhi,Qian:2021ihf,Chiesa:2021qpr,Liu:2021akf,Buttazzo:2021lzi,DiLuzio:2018jwd,Han:2020uak,Chen:2021pqi,Capdevilla:2021kcf,Cesarotti:2022ttv}. However, the advantages of colliding muons are accompained by the significant challenges related to designing a collider using unstable particles~\cite{sm21-imcc-accelerator}. The short lifetime of the muon requires an extremely fast operation from production, to acceleration, to the ultimate collision of the muon beams. The most mature muon production method relies on tertiary production of muons from protons striking a target~\cite{Delahaye:2013jla}. The emittance of the resulting muon beams must then be strongly reduced, for example via the “ionization cooling” process~\cite{MICE} before the acceleration stage. Alternative approaches~\cite{Alesini:2019tlf} to produce low emittance muon beams are being investigated, but still require substantial effort.

Finally, detectors and event reconstruction techniques~\cite{sm21-imcc-performance} need to be designed to cope with the presence of beam-induced background (BIB), a continuous flux of secondary and tertiary particles produced by the muon decays and their interaction with the machine elements. This paper reviews the available and incoming detector technologies that could be employed in an experiment at a muon collider, with special focus on a low-energy collider operating at $\sqrt{s}=3$~TeV, and outlines the main instrumentation challenges that will need to be addressed in building detectors able to successfully deliver the physics potential of a muon collider. It should be mentioned that development of these technologies is beneficial for other future collider options, in particular high energy hadron machines.

This article is organised as follows. Section~\ref{sec:requirements} briefly describes the expected beam structure, data-taking environment and the current proposed detector layout. Sections~\ref{sec:tracker} to~\ref{sec:muon} focus respectively on promising technologies that could be employed in the tracking detector, the calorimeter systems, and dedicated muon spectrometers. General considerations regarding trigger systems and data acquisition are discussed in Section~\ref{sec:tdaq}, while a summary and conclusions are presented in Section~\ref{sec:conclusions}.

\FloatBarrier

\section{Detection environment}
\label{sec:requirements}

\begin{figure}[h]
    \begin{center}
        \centering
        \includegraphics[width=0.5\textwidth]{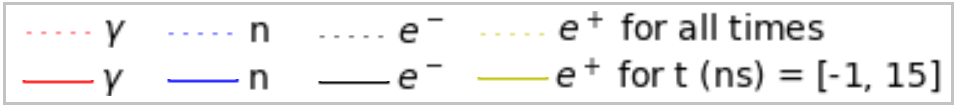}
        \includegraphics[width=\textwidth]{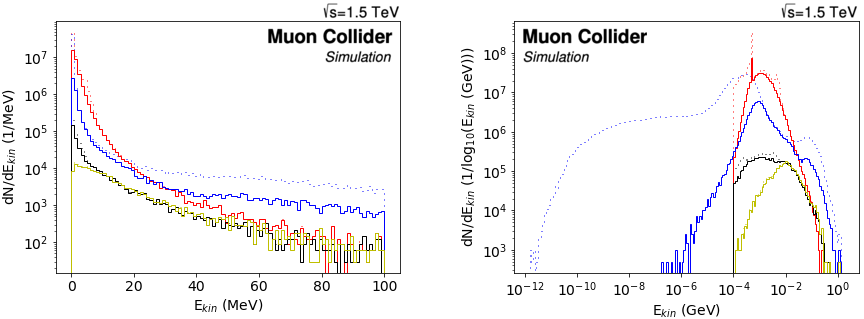}
    \end{center}
    \caption{
    Energy distribution (left) and lethargy plot (right) of BIB particles exiting the accelerator volume, divided by particle type, assuming $\sqrt{s} = 1.5$~TeV. Results by FLUKA for a $2\cdot10^{12}$~$\mu^-$ beam, decaying within 100~m from the IP. Energy cutoffs: 100 keV for $\gamma$, $e^-$, $e^+$ and 10$^{-14}$ GeV for neutrons. No time cut is applied to distributions represented by the dotted lines, while solid lines show only particles exiting within -1 and 15 ns.}
    \label{fig:bibspectra}
\end{figure}

The products composing the BIB create a large particle flux that interacts with the detector elements. This results in a unique experimental environment, with a related set of challenges, which sets muon colliders apart from the other collider facilities that are being considered for the future of high-energy physics.
The composition, flux, and energy spectra of the BIB making it into the detector depend strongly on the machine configuration of the interaction region, the Machine–Detector Interface (MDI), and the collision energy~\cite{Mokhov:2011zzd,Mokhov:2014hza,Bartosik:2019dzq,Lucchesi:2020dku}. 
The most important BIB property is that it is composed of low-energy particles, thanks to the MDI mitigation action, and is characterised by a broad arrival time in the detector. The expected spectra for the BIB particles entering the detector region in a collider operating at a centre-of-mass energy of $\sqrt{s} = 1.5$~TeV, as described by the FLUKA simulation of Ref.~\cite{Collamati:2021sbv}, are shown in Figure~\ref{fig:bibspectra}.

A full conceptual design of a multi-TeV muon collider facility has still to be performed. The overall goal would be to define a feasible end-to-end design of the collider complex and of the detectors, considering two interaction points (IP). This section briefly presents the main aspects of an envisaged experimental facility collecting data at a centre-of-mass energy of $\sqrt{s} = 3$~TeV for five years. Two single-bunch beams are produced and accelerated to be injected at 5~Hz into the collider ring.
Detailed studies to optimise the design of the interaction region to reduce BIB on detector to sustainable levels were performed by the MAP collaboration with the MARS15 Monte Carlo software~\cite{Mokhov:2017klc} and more recently with FLUKA~\cite{Ahdida:2022gjl,Battistoni:2015epi}. 
The background particle fluxes in the detector can be reduced by a substantial factor by two conical shaped shielding nozzles, described in Section~\ref{subsec:layout}. These nozzles, assisted by the magnetic field induced by a solenoidal magnet encasing the innermost detector region, could trap most of the electrons arising from muon decays close to the IP, as well as most of incoherent $e^{+}e^{-}$ pairs generated in the IP. With this sophisticated shielding in the MDI region, a total reduction of background loads by more than three orders of magnitude is obtained.
The design of the nozzles was optimized by MAP for the energy of $\sqrt{s} = 1.5$~TeV and implemented in the FLUKA simulation. The same shielding design was retained to study the background levels produced at $\sqrt{s} = 3$~TeV machine, without a dedicated MDI optimization. A first attempt to design the collider lattice of the interaction region at $\sqrt{s} = 10$~TeV adopting the same shielding design confirms that the BIB conditions expected at higher energies could be considered at the same level as at $\sqrt{s} = 1.5$~TeV for the following discussion.

The tentative target parameters for a muon collider at the different energies under study are summarised in Table~\ref{tab:beams} for $\sqrt{s} = 1.5$~TeV, $\sqrt{s} = 3$~TeV and $\sqrt{s} = 10$~TeV.
\begin{table}[!h]
    \footnotesize
    \centering
    \caption{Muon collider parameters, assuming accelerator complexes operating at $\sqrt{s} = 1.5$, 3 and 10~TeV.}
    \begin{tabular}{lccc}
        \toprule
        Parameter                                                      & $\sqrt{s} = 1.5$~TeV & $\sqrt{s} = 3$~TeV  &
        $\sqrt{s} = 10$~TeV \\
        \midrule
        Beam momentum [GeV]                                            & 750                  & 1500                & 5000 \\
        Beam momentum spread [\%]                                      & 0.1                  & 0.1                 & 0.1 \\
        Bunch intensity                                                & $2 \cdot 10^{12}$   & $2.2 \cdot 10^{12}$ & $1.8 \cdot 10^{12}$ \\
        $\beta^{*}_{x,y}$ [cm]                                         & 1                    & 0.5                 & 0.15\\
        $\epsilon_{TN}$ normalised transverse emittance [$\pi \,\mu$m rad]   & 25             & 25                  & 25 \\
        $\epsilon_{LN}$ normalised longitudinal emittance [MeV m]      & 7.5                  & 7.5                 & 7.5 \\
        $\sigma_{x,y}$ beam size [$\mu$m]                              & 6                    & 3                   & 0.9  \\
        $\sigma_{z}$ beam size [mm]                                    & 10                   & 5                   & 1.5 \\
        \bottomrule
    \end{tabular}
    \label{tab:beams}
\end{table}

\subsection{Detector layout}
\label{subsec:layout}

\begin{figure}[h]
    \begin{center}
        \centering
        \includegraphics[trim={0.8cm 0.8cm 0.8cm 0.8cm},clip,width=0.6\textwidth]{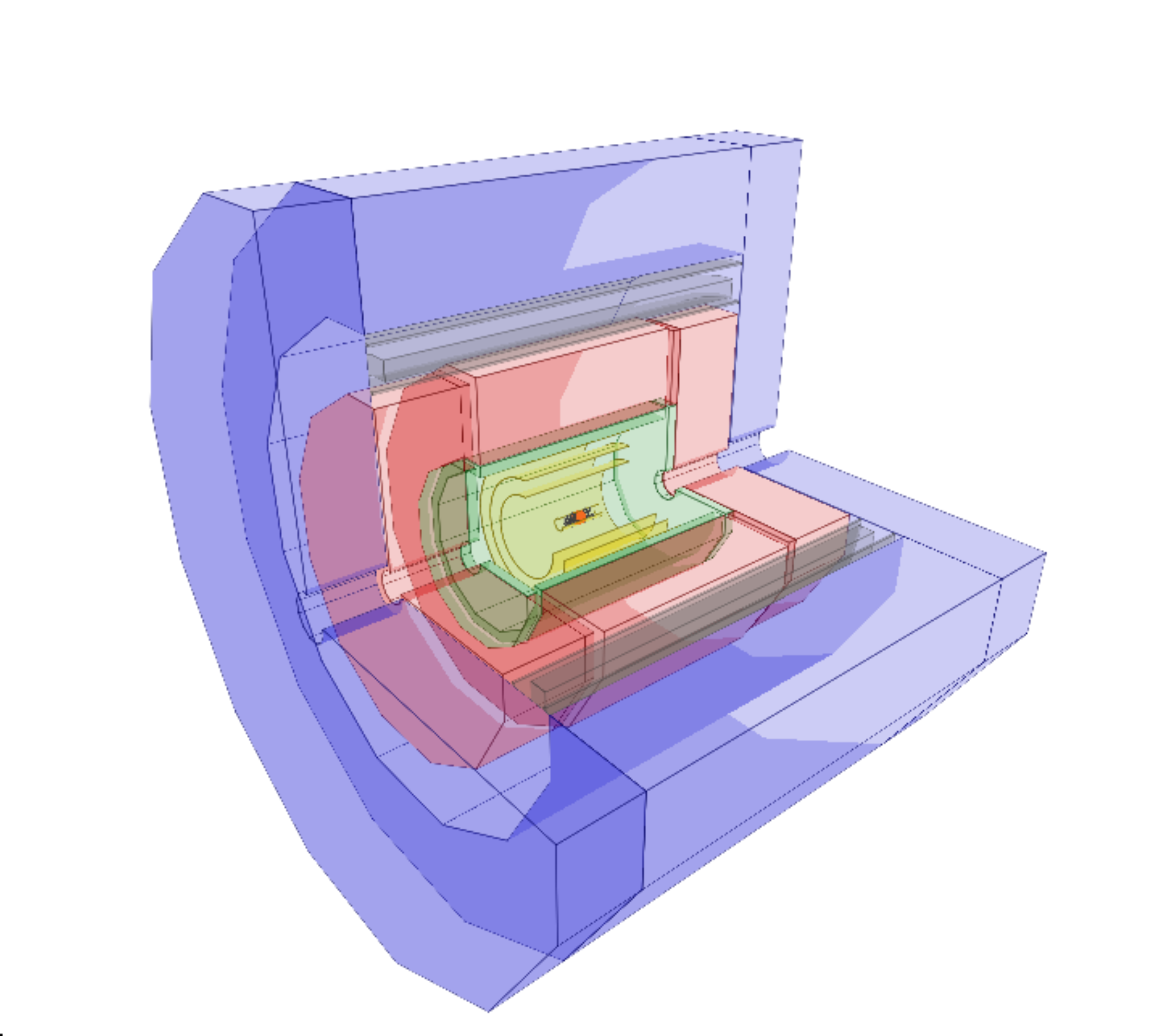}
    \end{center}
    \caption{Illustration of the full detector, from the \GEANT 4 model. Different colours represent different sub-detector systems: the innermost region, highlighted in the yellow shade, represents the tracking detectors. The green and red elements represent the calorimeter system, while the blue outermost shell represents the magnet return yoke instrumented with muon chambers. The space between the calorimeters and the return yoke is occupied by a 3.57~T solenoid magnet.}
    \label{fig:detector}
\end{figure}

The design of a dedicated experiment is still in its infancy, but some general considerations already hold true. Given the breadth of the expected physics programme, a hermetic detector with angular coverage as close as possible to $4\pi$ is required.
The detector will be based on a cylindrical layout and will include an inner tracking detector immersed in a magnetic field, a set of calorimeter systems designed to fully contain the products of the muon collisions, and an external muon spectrometer. 
The most recent detector model is based on the CLICdet concept~\cite{CLICdp:2017vju,CLICdp:2018vnx,ILDConceptGroup:2020sfq,ILC:2007vrf} with relevant modifications to the tracker topology,  introduced to mitigate the impact of the BIB: two double–cone shielding absorbers made of tungsten with a borated polyethylene (BCH2) coating,  are located inside the detector in the forward regions\footnote{A right-handed coordinate system with its origin at the nominal interaction point in the centre of the detector is used. Cylindrical coordinates $(r,\phi)$ are used in the transverse plane, $\phi$ being the azimuthal angle around the z-axis and $\theta$ the polar angle with respect to the z-axis.} along the beam axis at $\theta < 10^{\circ}$ in the 6 to 600~cm region from the IP.
The innermost system consists of a full-silicon tracking detector. The tracking detector is surrounded by a calorimeter system formed by an electromagnetic calorimeter (ECAL) and a hadronic calorimeter (HCAL), and is immersed in a magnetic field of 3.57~T provided by a solenoid with an inner bore of 3.5~m. Finally, the outermost part of the detector consists of a magnet yoke designed to contain the return flux of the magnetic field and is instrumented with muon chambers. The full detector is shown in Figure~\ref{fig:detector}. 

\subsection{Radiation levels}

The simulation of the BIB distributions and rates is a crucial element to quantify the requirements on the detector components. A preliminary estimate from FLUKA simulations of the $\sqrt{s} = 1.5$~TeV conditions, and that does not include the contribution from the muon collisions, was recently released in Ref.~\cite{Collamati:2021sbv}. The simulation employed a simplified detector geometry: all the silicon layers composing the inner tracker were included with exact dimensions. The calorimeters, magnetic coils, and the return yoke were approximated with cylindrical elements with densities and material composition based on the averages from the full geometry. The magnetic fields were assumed to be uniform.

\begin{figure}[h]
    \begin{center}
        \centering
        \includegraphics[width=.8\textwidth]{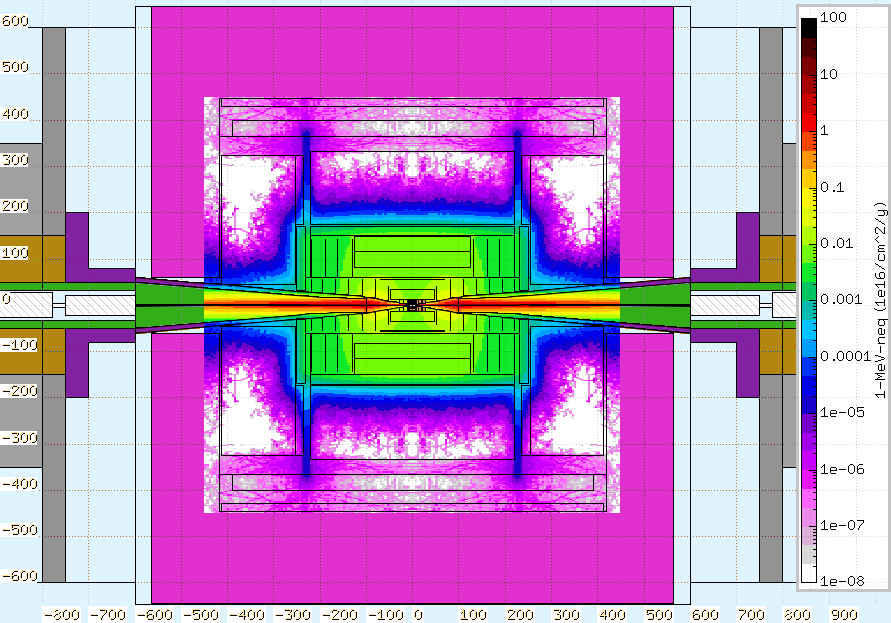}
    \end{center}
    \caption{Map of the 1-MeV-neq fluence in the detector region for a muon collider operating at $\sqrt{s} = 1.5$~TeV with the parameters reported in Table~\ref{tab:beams}, shown as a function of the position along the beam axis and the radius. The map is normalised to one year of operation (200 days/year) for a 2.5 km circumference ring with 5 Hz injection frequency.}
    \label{fig:fluence}
\end{figure}

\begin{figure}[h]
    \begin{center}
        \centering
        \includegraphics[width=.8\textwidth]{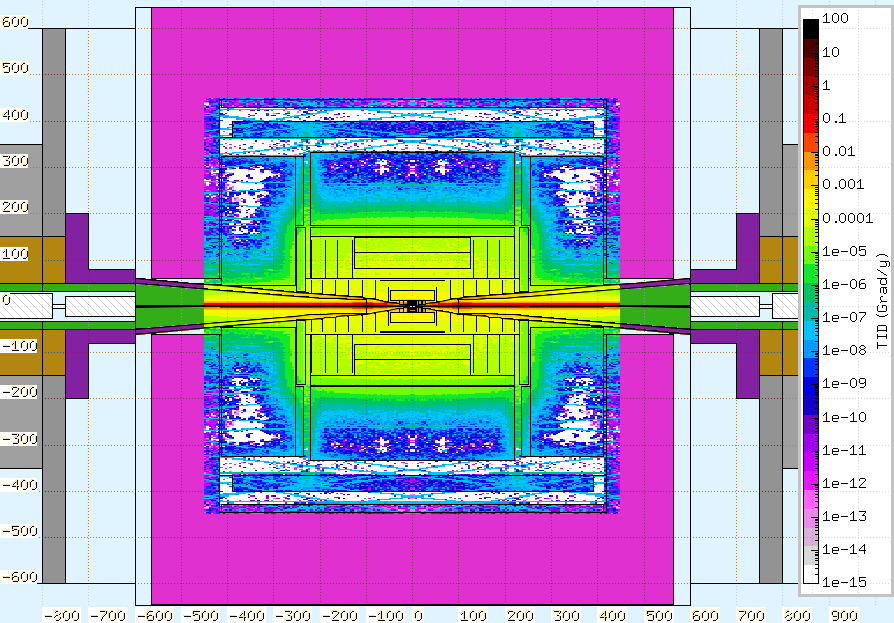}
    \end{center}
    \caption{Map of the TID in the detector region for a muon collider operating at $\sqrt{s} = 1.5$~TeV with the parameters reported in Table~\ref{tab:beams}, shown as a function of the position along the beam axis and the radius. The map is normalised to one year of operation (200 days/year) for a 2.5 km circumference ring with 5~Hz injection frequency.}
    \label{fig:tid}
\end{figure}

Figures~\ref{fig:fluence} and~\ref{fig:tid} show respectively the expected 1 MeV neutron equivalent fluence (1-MeV-neq) and the total ionizing dose (TID) in the detector region, shown as a function of the beam axis z and the radial distance $r$ from the beam axis. The normalization for the dose maps is computed considering a 2.5~km circumference ring and an injection frequency of 5~Hz. 
Assuming 200 days of operation during a year, the 1-MeV-neq fluence is expected to be $\sim 10^{14-15}$~cm$^{-2}$y$^{-1}$ in the region of the tracking detector and of $\sim 10^{14}$~cm$^{-2}$y$^{-1}$ in the electromagnetic calorimeter, with a steeply decreasing radial dependence beyond it. The total ionizing dose is $\sim10^{-3}$~Grad/y on the tracking system and $\sim10^{-4}$~Grad/y on the electromagnetic calorimeter.

\FloatBarrier

\section{Tracking systems}
\label{sec:tracker}

The ability to reconstruct trajectories of charged particles in the tracking system and to measure their parameters with high precision is essential at the muon collider experiments. 
In the expected operating conditions, high performance tracking is necessary to achieve good efficiency and resolution for reconstructing charged leptons, jets, energy sums, displaced vertices originating from the heavy flavor jets, as well as potential new phenomena. 

\begin{figure}[!ht]
\center
\includegraphics[width=0.7\textwidth]{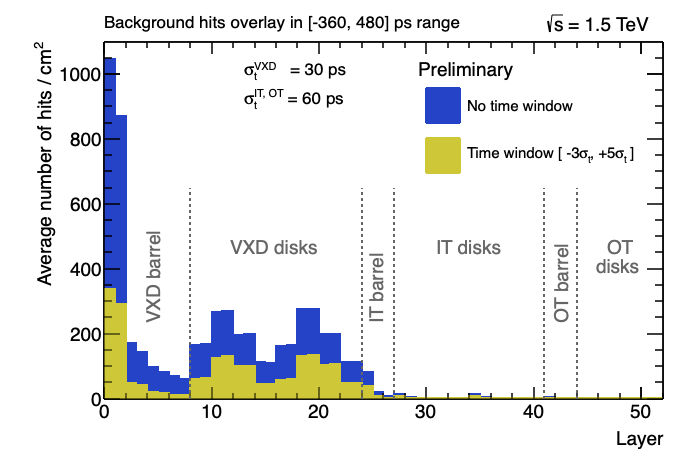}
\caption{Average hit density per bunch crossing in the tracker and as a function of the detector layer.}
\label{fig:HitMultTracker}
\end{figure}

The BIB represents a significant hurdle for tracking, both in terms of the data volumes generated by the tracker as well as by introducing a combinatorial challenge in the track reconstruction. In each bunch crossing, BIB particles on average generate 500,000 hits in the most inner layer of the tracker, located just few centimeters  away from the interaction point. This corresponds to hit density of up to 1,000 hits/cm$^2$; the density does however decrease rapidly as a function of the radial distance from the beam-line, as shown in Figure~\ref{fig:HitMultTracker}. 

It is clear from these numbers that high granularity of silicon pixels is necessary in order to achieve hit occupancy level of a few \%. In addition, various handles to reduce the BIB should be explored for both on- and off-detector filtering. Possible filtering schemes include:
\begin{itemize}
\item \textbf{Timing}: Removing out of time hits appears to reduce the data load by a factor of 3. Timing information will eventually be needed in the reconstruction, but it makes sense to apply initial filtering on-detector.
\item \textbf{Clustering}: Clustering reduces the number of pixel groups read out. This requires more on-detector processing and results in more bits per cluster and a higher power budget, but can reduce the number of hits read out. Selection requirements can also be applied to the cluster shape. The effectiveness needs to be assessed for each BIB cluster type.
\item \textbf{Energy Deposition}: Each of the backgrounds has a characteristic energy deposition signature. For example neutrons have low, localized energy 
deposit. This should be a useful requirement to make on-detector.
\item  \textbf{Correlation Between Layers}: This is a powerful handle for background rejection. However, implementation may be complex and costly, doubling the number of channels. For on-detector filtering, it also requires transfer of data between layers in a very busy environment.
\item \textbf{Local Track Angle}: Track angle measurement can be made in a single detector if the thickness/pitch ratio distributes the signal over several pixels. This avoids the complexity of inter-detector connections and could provide a monolithic solution~\cite{Lipton:2022njd,Lipton:2019drv}.
\item \textbf{Pulse Shape}: Signals from BIB can come with a variety of angles and may not give the deposit profile and pulse shape of a typical MIP.  Appropriate pulse processing, such as multiple sampling, RC-CR filters, zero crossing, or delay line clipping can be used to further reduce the data load.
\end{itemize}

The basic trade-offs are between the complexity, power, and mass needed to implement a on-detector filter, and the benefit of reduced data rate. One has to be careful however, in particular when it comes to on-detector filtering. Overly aggressive front-end filtering schemes can introduce irrecoverable inefficiencies and biases in track reconstruction and can limit acceptance for some new physics signatures, such as those of long-lived particles. 

A recent simulation study was conducted to determine granularity and timing requirements for the tracker sensors in order to reduce the hit occupancy to under 1\% target level. In this study, we varied pixel size and per hit timing resolution in each layer of the detector. Hits were integrated in the time period of 1~ns following the bunch crossing. It was found that for the vertex detector granularity of $25 \times 25$~$\mu m^{2}$ and time resolution of 30~ps were needed to achieve the desired occupancy goal. In the inner tracker, asymmetric macropixels with $50 \times 100$~$\mu m^{2}$ size and 60 ps timing resolution were sufficient to satisfy the requirement, while in the outer tracker featured microstrips with the size of 50~$\mu m \times 10$~mm and 60~ps time resolution. It is thus evident that R\&D efforts towards 4D tracking are necessary to achieve the required spacial granularity and timing resolution. Below we provide a description of promising technologies for achieving such goals. 

\subsection{Silicon Sensor Technology}
Silicon-based sensors have come to dominate the technology for collider detector tracking systems. This is likely to continue into the muon collider era. In the past decade there have been a number of technological developments that promise to achieve many of the capabilities discussed in the previous section. They address different aspects of the needs for space and time resolution, pattern recognition, electronics integration, radiation hardness, and low cost. We list some candidate technologies along with associated R\&D areas:

\subsubsection{Monolithic Devices (CMOS MAPS)}
\begin{figure}[h]
\centering
\includegraphics[width=0.95\textwidth]{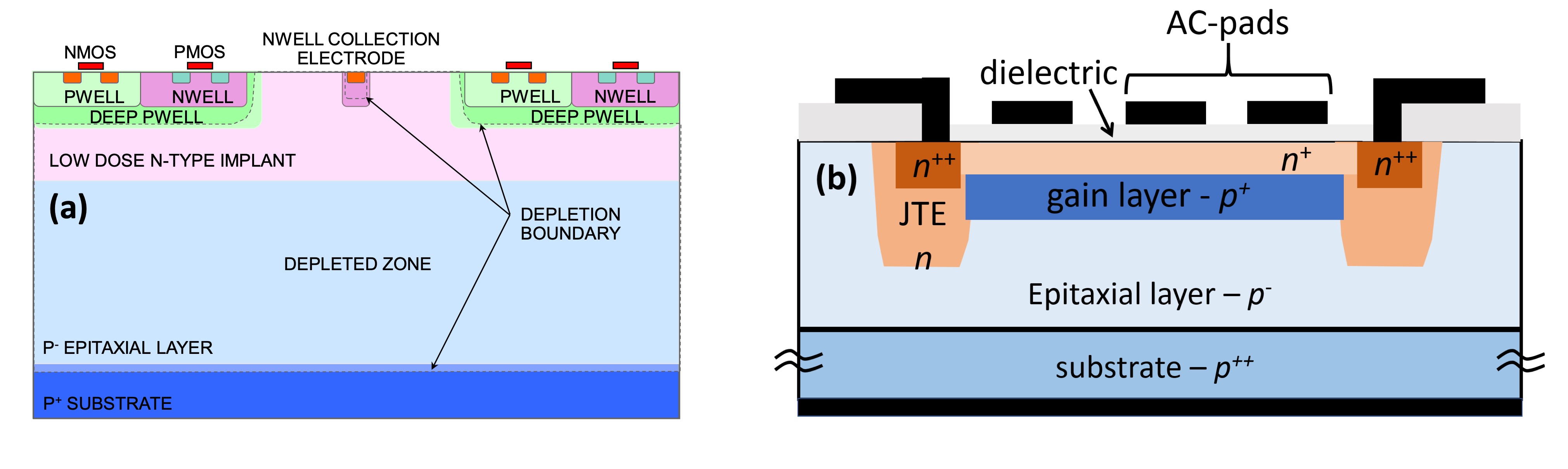}
\caption{a) Cross section of the MALTA~\cite{Cardella:2019ksc} CMOS sensor showing the implant structure.  The low dose n-implant provides an electrode that improves the uniformity of the p-epitaxial drift field.  b) Cross section of an AC coupled LGAD~\cite{Giacomini:2019kqz}.  The Junction Termination Edge (JTE in the figure), which would normally separate each pixel, is only needed at the edge of the device, providing near 100\% fill factor.}
\label{CMOS_LGAD}
\end{figure}

CMOS Monolithic Active Pixel Sensors (MAPS) are based on standard CMOS process flows with thick (20--50~$\mu$m) epitaxial layers. Charge is collected from electron-hole pairs generated in the epitaxy. There is a small area n-type collection electrode with the CMOS circuitry embedded in a deep $p$-well to avoid parasitic charge collection by the CMOS transistors. The geometry of the electrodes and associated circuitry means that the epitaxy is difficult to deplete evenly. The first such devices used diffusion rather than drift to collect charge.  Recent prototypes (shown in Figure~\ref{CMOS_LGAD}a) have added a deep, lightly doped, n-layer below the p-well to provide a more uniform drift field in the epitaxy~\cite{Cardella:2019ksc}. 

The proximity of the CMOS transistors to the epitaxy means these devices have significant sensitivity to analog-digital crosstalk. It is likely that the CMOS analog section would have to be 3D stacked with digital TDC, ADC and I/O tiers to achieve adequate isolation and overall functionality.  Radiation hardness needs to be studied and improved.  In particular the effects of doping evolution in the epitaxy will affect fields and operation as the device ages. This is 
a particular problem with the large ratio of collection node to collection area which forces a difficult geometry for the collection field. The size of CMOS sensors is limited by the typically 2x3 cm CMOS reticule. To achieve a large area device the reticules must be "stitched" using an additional metal layer or tiled while minimizing dead area at the edges.

\subsubsection{Devices with Intrinsic Gain}
In sensors based on the Low Gain Avalanche Diode (LGAD) design, the initial charge created by an impinging particle is amplified in a ``gain layer'' by a factor of about 10--30. The resulting current signal is large and fast, enabling a 20--30~ps time resolution. An interesting feature of LGAD-based sensors is that the associated front-end requires less power since the sensor provides the first amplification stage. 
The LGAD  design is relatively new and is undergoing rapid development.
The design of the current generation of LGADs being used for endcap timing layers in the CMS and ATLAS HL-LHC upgrades has now been superseded. This first generation of devices suffers from limited fill-factor due to edge field limiting structures and moderate radiation hardness. Recent works have combined internal gain with the novel resistive read-out design, reaching a design (the so-called RSD, Resistive Silicon Detectors) with a 100\% fill factor and excellent spatial precision even with pixels with a large pitch (a precision of less than 5\% of the pitch)~\cite{Tornago:2020otn}. Parallel to this development, new studies suggest possible improvements in the radiation resistance of the LGAD design. The two most promising are: (i) the insertion of an additional layer of carbon to reduce acceptor removal in the gain implant and (ii) the design of the gain implant using both acceptor and donor dopings so that acceptor removal is compensated by acceptor removal, extending the radiation hardness of the gain implant. 
The resistive charge-sharing design also allows for sparser read-out geometries, limiting the density of analog channels. 

For the muon collider,  being a high-rate environment, the novel DC-coupled design of the resistive read-out technique might be very beneficial, as it allows for a faster recovery time. 
Detailed studies of AC-RSD and DC-RSD will be needed, optimizing the electrode density and geometry in conjunction with charge deposition characteristics of the photon, electron/positron, and neutron backgrounds. Dopant removal is the primary limitation to radiation hardness, and continued study is needed to understand the practical restrictions imposed by this effect. The total current going into detector bias will be larger in devices with gain and may represent a significant fraction of the total power.


\subsubsection{Hybrid Small Pixel devices}
\begin{figure}[h]
\centering
\includegraphics[width=0.95\textwidth]{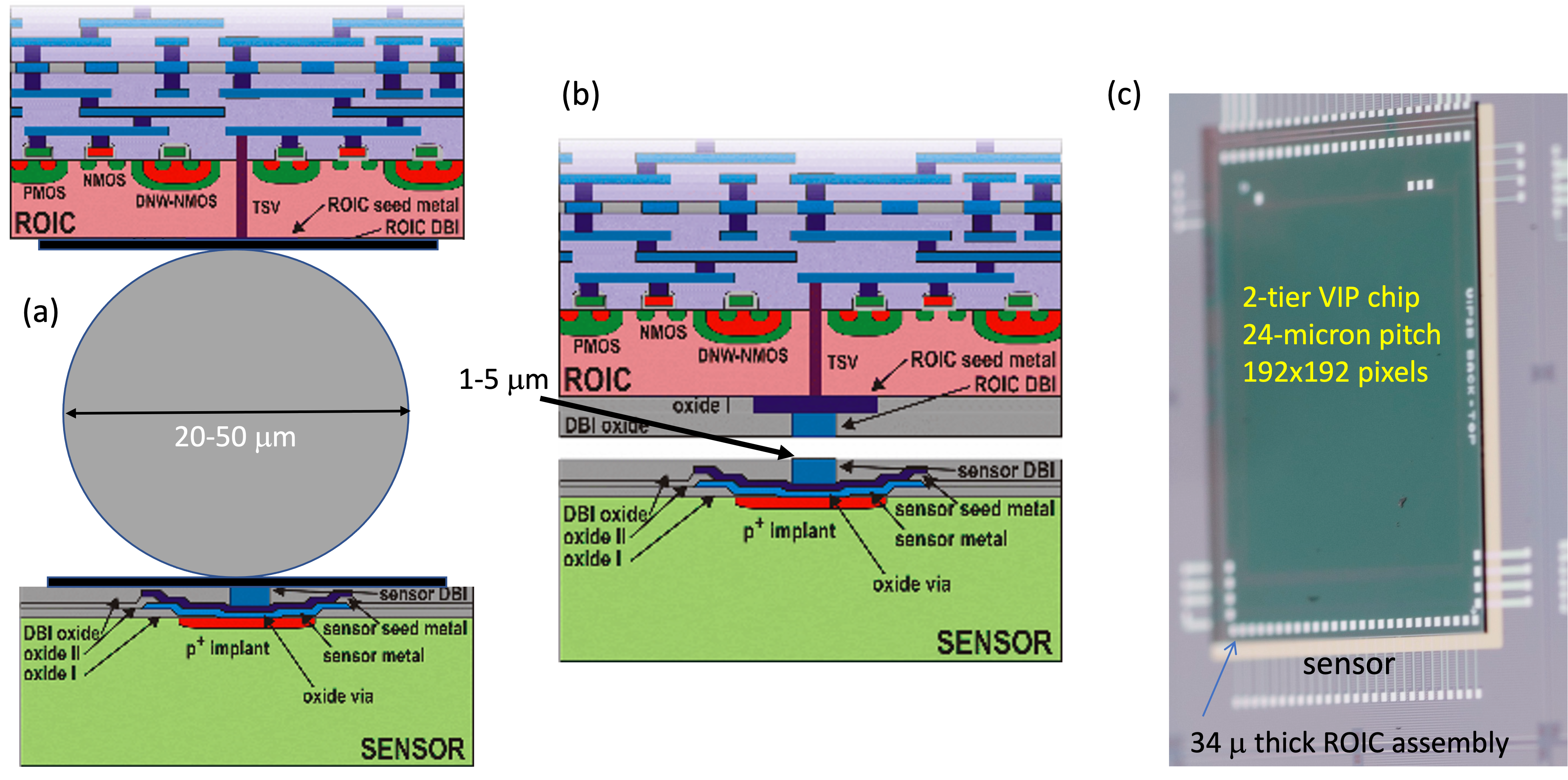}
\caption{a) A typical bump bonded sensor/readout chip geometry. The spacing is determined by the size of the bump and under-bump metalization pad.  b) A hybrid bonded sensor/readout stack with the pitch limited by the micron-level hybrid metalization imbedded in the top oxide layer.  c) An example of a three-tier hybrid bonded stack with separate analog and digital readout layers\cite{Lipton:2015vca}. The readout pitch is 24 microns and the readout stack thickness is 35$\mu$}
\label{fig:Bump_Hybrid}
\end{figure}

These are standard "hybrid" pixel diode sensors without gain. However fast timing and good position resolution can be achieved with fine pixel pitch and low input capacitance. Bump bonding using solder, indium, or copper, is limited to $\approx$15-50 micron interconnect pitch. Three-dimensional hybrid bonding, illustrated in Figure~\ref{fig:Bump_Hybrid},\footnote{The electronics industry has adopted "hybrid bonding" as the generic term for oxide bonded stacks of wafers or chips with imbedded metal to achieve top to bottom electrical connection. This technology is commonly used in cell phone image sensors.} can achieve both $<5~\mu$m pitch and low enough interconnect capacitance to meet noise and power limitations. The sensors have the advantage of being intrinsically radiation hard with signal/noise large enough to provide 20-30 ps time resolution.  It will likely be necessary to use a design where ADCs and TDC service multiple small pixels to reduce the density of ADCs and TDCs, thus reducing both the in-pixel density and power. 

\subsubsection{Intelligent Sensors}
The different characteristic signals generated by the electromagnetic, neutron, and charged hadron backgrounds and signal MIPs prompts consideration of more "intelligent" sensors that can separate the BIB from the signal. An example is the current 2-layer track trigger design for CMS at the LHC where low $p_{\textrm{T}}$ tracks are filtered out by comparing hits on separated sensor layers. Such multi-layer designs are limited by the complex interconnection and data transmission paths needed to communicate between sensor layers. However for a device where the thickness/pixel pitch ratio is large enough the pixel pulse shapes and cluster patterns will be very different for MIPs and BIB hits. This information can be utilized for a prompt local filter to reject BIB. Radiation-induced traps will cause the pulse shapes and induced current patterns to change during the lifetime of the detector, possibly necessitating changes in algorithms.

Appropriate information density be achieved in small pixel devices or double-sided LGADs~\cite{Lipton:2022njd}. In the double sided LGAD fast timing signals are read on a top, larger pitch layer coupled to the gain layer. Charge deposition patterns and timing are reconstructed on a bottom, pixelated layer. Other concepts can be explored where the very different pulse shapes and patterns can be used to separate signal from BIB, perhaps incorporating on-chip ML techniques.


\subsection{Power Considerations}
This section describes an attempt to estimate the power constraints on the tracker based on extrapolations of the existing technologies. The study focuses on the vertex detector and assumes a design with 25$~\mu$m$^2$ pixels with four barrel layers and four endcap disks on each side, as described earlier in this paper. Conventional scaled CMOS electronics~\cite{OConnor:1999isd} and possible extrapolations of optical-based data transmission are also assumed. New technologies might change the picture completely. 

For conventional CMOS-based amplifiers a conventional silicon detector operating with no internal gain the front-end power will be determined by the capacitive load on the front-end and the desired signal/noise and rise time. For example a simple SPICE model of a preamplifier loaded with 100~fF capacitance provides 4~ps time jitter for 1 $\mu$A bias current and  45~ps for a bias current of 100~nA~\cite{RLDDesign}. A time jitter of $< 30$~ps can be achieved in a conventional sensor with 50~$\mu$m thickness and front-end current of less than 250~pA if the detector capacitance is carefully controlled using 3D interconnections. If the sensor is based on LGAD-like internal gain the signal presented to the preamplifier can be $10-20$ times larger. For the same signal over noise, this could reduce the front-end transductance and associated drain current by the roughly the square root of the gain. For a conventional CMOS amplifier we can expect about 450~W of power into the vertex detector for analog bias.

In addition to front-end power there is the power necessary to bias the detector. This can become significant for heavily irradiated detectors. If an HL-LHC-like operating scenario is considered, the final depletion voltage can be as high as  500~V (depending on the technology chosen). Under these conditions the vertex sensor bias power is about 100~W.

It is also useful to estimate the detector data load. Using the simulated layer occupancies in the vertex detector, we obtain a total rate of hits of $15 \times 10^{13}$ bits per second (b/s) for the vertex detector. More detail about the estimate is provided in Section~\ref{sec:tdaq}. The power needed per bit for the Low Power Gigabit Transceiver (lpGBT)~\cite{lpGBT} is about 41~pJ/bit. Power efficient optical transmission is the subject of intense study by the semiconductor industry and we take 10~pJoule/bit as a conservative estimate for future power consumption. This gives us a data transmission power of about 1.5~kW. AttoJoule/bit levels, albeit before radiation damage considerations, appear feasible in the near future~\cite{7805240}, further reducing the penalty for reading the full event. Each link must be capable of transmitting at a rate of about $20$~Gb/s to limit the number of physical optical connections to one per module. Higher speeds (if available) will lead to a reduction of the overall number of optical links in the system. 
\FloatBarrier

\section{Calorimeter systems}
\label{sec:calorimeter}

The measurement of physics processes at the energy frontier requires excellent energy and spatial resolutions to resolve the structure of collimated high-energy jets. Jet reconstruction, and the improvement of the reconstructed jet energy resolution, is the driving theme of ongoing R\&D activities in high energy physics, and a muon collider will not diverge from this general theme. Future lepton colliders aim at separating W and Z bosons in the dijet channel, which requires a 3-4\% jet energy resolution for jets above 100~GeV.

\begin{figure}[!ht]
     \center
        \includegraphics[width=0.48\textwidth]{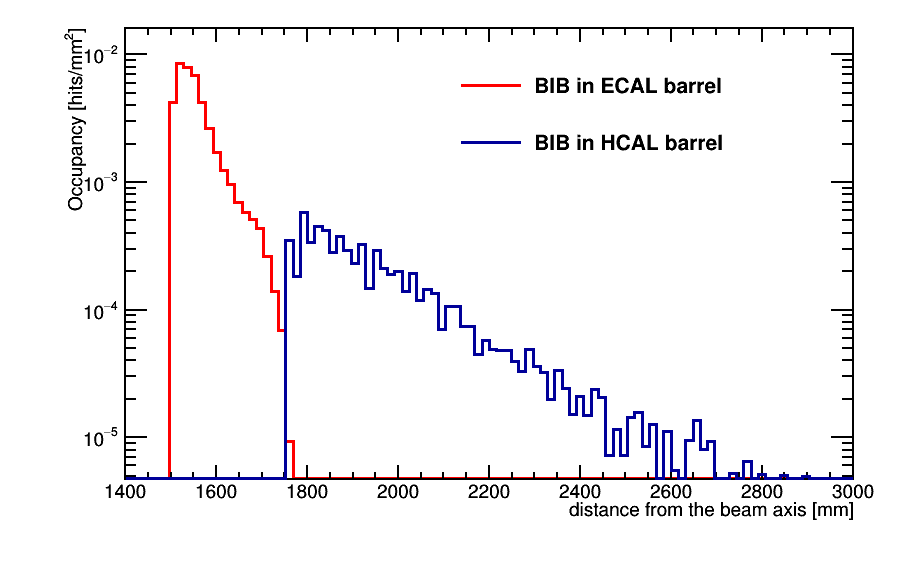}        \includegraphics[width=0.48\textwidth]{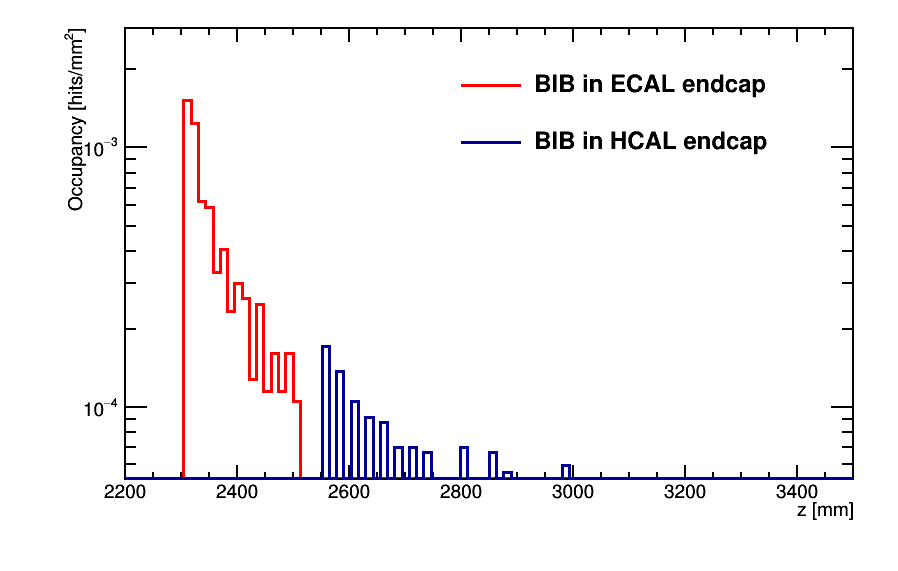}
        \caption{Hit occupancy in the calorimeter barrels (left) and endcaps (right) in a single bunch-crossing at $\sqrt{s}=1.5$ TeV.}
        \label{fig:occupancy}
   \end{figure}

In a multi-TeV muon collider, the calorimeter system has to operate in the intense flux of low energy particles arising from the BIB. The BIB in the calorimeter region is mainly formed by photons (96\%) and neutrons (4\%). In the current detector layout, a flux of about 300 particles per cm$^2$ is present at the ECAL barrel surface, with an average photon energy of about 1.7~MeV. Given the high flux, several particles may overlap in a single cell, resulting in a hit where their energy is summed up. The occupancy, defined as the number of hits per mm$^2$ in a calorimeter layer, as a function of the calorimeter depth is shown in Figure~\ref{fig:occupancy}. The simulation shows how the ECAL system is expected to absorb most of the BIB radiation, resulting in a significantly lower occupancy for the HCAL system.
The ECAL is expected to receive approximately 100~krad/y of total ionising dose and a $10^{13-14}$~cm$^{-2}$~1-MeV-neq fluence. 

\begin{figure}[!ht]
     \center
        \includegraphics[width=0.48\textwidth]{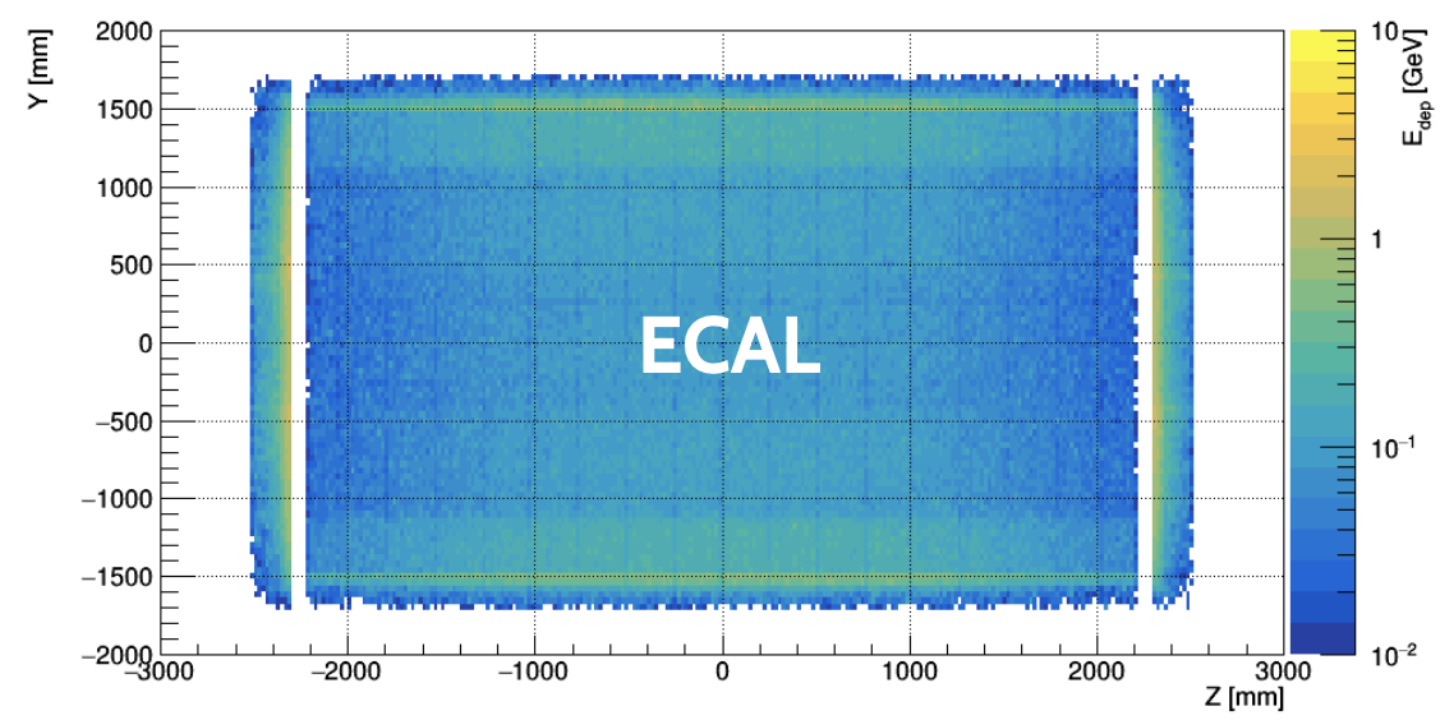}        \includegraphics[width=0.48\textwidth]{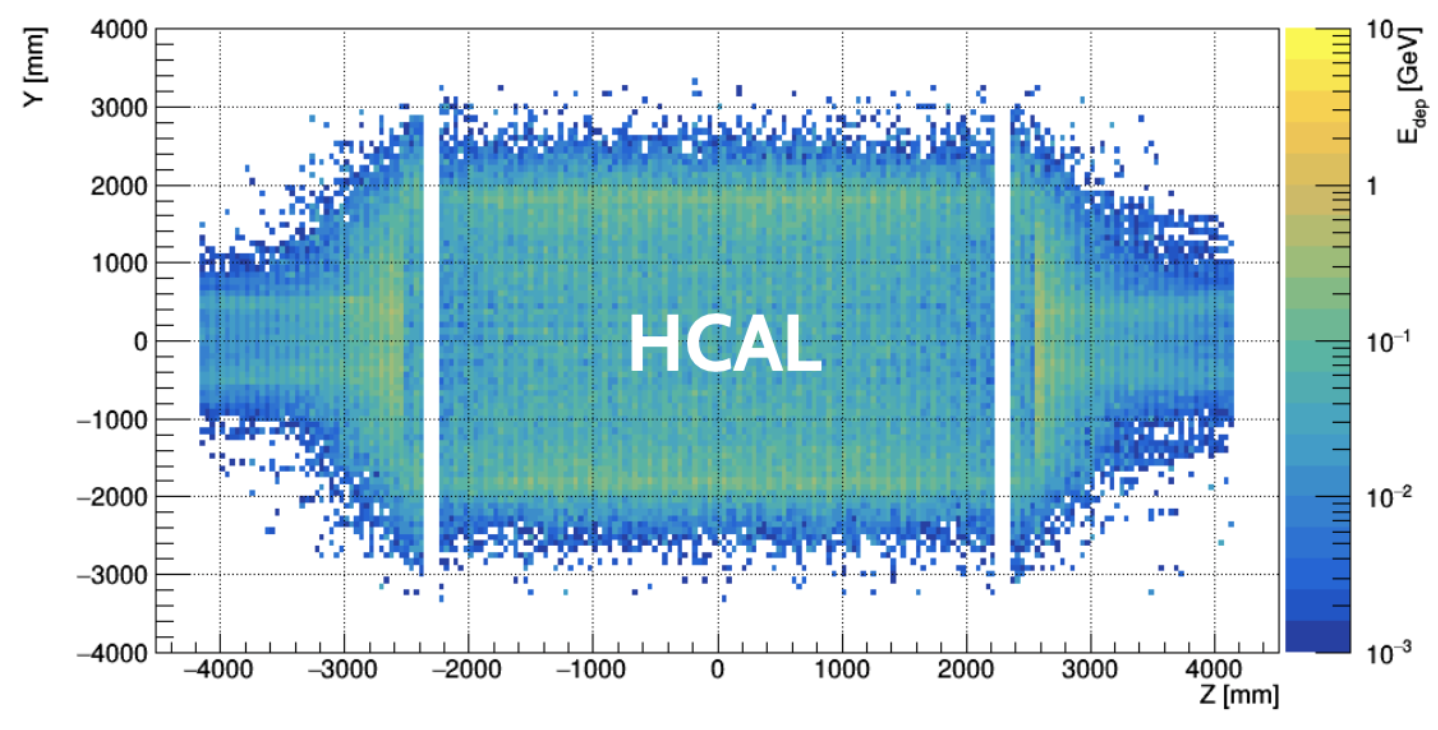}
        \caption{Energy deposited by the BIB in a single bunch-crossing at $\sqrt{s}=1.5$ TeV, in ECAL (left) and in HCAL (right).}
        \label{fig:edeposit}
   \end{figure}

The spatial distribution of the energy deposited in ECAL and HCAL in a single bunch-crossing is shown in Figure~\ref{fig:edeposit}. 
   
Similarly to what can be done in the tracker, the time of arrival of particles in a calorimeter cell can be exploited to discriminate the BIB contributions from the primary interactions. At the same time, the longitudinal energy deposit distribution from BIB particles is expected to be deposited in the innermost layers of the calorimeter, with particles from the primary interaction propagating deeper in the detector.

The technology and the design of the calorimeters should be chosen to reduce the effect of the BIB, while keeping good physics performance. Several requirements can be inferred:
\begin{itemize}
    \item \textbf{High granularity} to reduce the overlap of BIB particles in the same calorimeter cell. The overlap can produce hits with an energy similar to the signal, making harder to distinguish it from the BIB;
    \item \textbf{Good timing} to reduce the out-of-time component of the BIB. An acquisition time window of about $\Delta t = 300$~ps could be applied to remove most of the BIB, while preserving most of the signal. This means that a time resolution in the order of $\sigma_t = 100$~ps (from $\Delta t \approx 3 \sigma_t$) should be achieved;
    \item \textbf{Longitudinal segmentation}: the energy profile in the longitudinal direction is different between the signal and the BIB, hence a segmentation of the calorimeter can help in distinguishing the signal showers from the fake showers produces by the BIB;
    \item \textbf{Good energy resolution} of $\frac{10\%}{\sqrt{E}}$ in the ECAL system is expected to be enough to obtain good physics performance, as has been already demonstrated for conceptual particle flow calorimeters.
\end{itemize}

The requirements imposed by the need to house the calorimeter systems within a large magnetic coil tend to disfavour designs fully based on homogeneous calorimetry. Sampling calorimeters based on alternating dense passive materials, such as copper, steel, or tungsten, and active readout materials, such as plastic scintillators, silicon, or gaseous detectors are likely to be employed, at least in the HCAL.
Two major approaches are being pursued to exploit sampling calorimeters and improve upon the current generation of collider experiments: multi-readout (dual or triple)~\cite{Lee:2017shn,dr20,b} and particle flow~\cite{Thomson:2009rp} calorimetry.
The first approach focuses on reducing the fluctuations in the hadronic shower reconstruction, which are the main responsible for the deterioration in the determination of the jet energy. This goal is achieved by measuring independently the electromagnetic and the non-electromagnetic components of a hadronic shower, thus allowing to correct event-by-event for the different response of the calorimeter to various particle species. 
The second approach focuses on the reconstruction of the four-momenta of every particle recorded by the detector. This method exploits tracking information and requires a detector with extreme granularity, combined with powerful reconstruction algorithms aimed at resolving each particle's trajectory through the whole detector.

\subsection{Dual-readout calorimetry}
The energy resolution of a calorimeter system is affected by fluctuations in the energy deposited in its active elements. When measuring hadronic showers, the fluctuations in the electromagnetic component of a shower represent the dominant contribution to the total resolution.
This effect can be mitigated by designing a calorimeter with the same response to electromagnetic (e.m.) and non-e.m. energy deposits (i.e. a compensating calorimeter with e/h=1.0), or by separately measuring the e.m. and hadronic energy deposits on an event-by-event basis.
The latter solution can be realized by measuring both the scintillation light component (sensitive to both hadrons and e.m. particles) and the Cherenkov light component (sensitive only to relativistic e.m. particles) and is called Dual-Readout calorimetry. Notable examples of implementations of this concept can be found in the work of the DREAM/RD52 collaboration~\cite{Wigmans:2007es} and the proposed IDEA~\cite{FCC:2018evy, CEPCStudyGroup:2018rmc} detector for FCCee/CepC.
In these calorimeters, signals are generated in scintillating fibers, which measure the deposited energy, and in clear plastic PMMA or quartz fibers, which are sensitive to the Cherenkov light. A large number of such fibers are embedded in a fully projective lead or copper absorber structure. This detector is not longitudinally segmented: e.m. and hadronic showers can be distinguished using short and long fibers in the calorimeter, and longitudinal information could be extracted by reading out the fibers on both ends. Past results~\cite{Lee:2017xss,Lee:2017shn} have demonstrated to reach an energy resolution for charged pions of 34\%/$\sqrt{\textrm{GeV}}$ and the final goal of 30\%/$\sqrt{\textrm{GeV}}$ seems well within reach. The main challenges appear in the handling of the high number of fibers and SiPMs which are of the order of few 10$^8$ and constitute an important fraction of the technology cost. 
The implementation of a third fiber material, or alternatively the time readout of the scintillation fibers, can be used to measure the MeV-scale neutrons produced in a hadronic shower and suppress the fluctuations arising from binding energy loss~\cite{Lee:2017oye}. This latter method is referred to as triple readout.
The absence of longitudinal segmentation is likely the limiting factor for deploying such design in a muon collider detector. However, hybrid designs composed of an ECAL made of crystals (such as LYSO or PbWO$_{4}$) and the hadronic section based on the DREAM fiber prototype have been also proposed~\cite{Lucchini:2020bac,Gaudio:2011zzb}. These designs need to demonstrate the feasibility of dual readout concepts in the crystal matrix.

\subsection{Particle flow calorimetry}
In recent years the concept of high granularity particle flow calorimetry~\cite{Brient:2001fow} has been developed in the context of the proposed International Linear Collider (ILC). The CALICE collaboration is the major developer of calorimeter concepts and technologies for highly granular detectors for particle flow. The goal of particle flow calorimeters is to build an image of the showers induced by the various jet-fragments to allow the correct matching of these showers with the charged particles measured in the tracker. This in turn enables to correctly identify and measure the energy of the showers induced by neutral hadrons.

\subsubsection{Silicon-based sampling}
With a radiation length ($X_0$) of 0.35~cm and an interaction length ($\lambda_I$) of 9.95~cm, tungsten is an ideal absorber for an electromagnetic sampling calorimeter. Silicon sensors can be used as active elements to achieve a high channel granularity and longitudinal segmentation. Moreover, state-of-the-art silicon sensors can sustain the high radiation dose of the expected BIB. Analogous technologies are adopted by LHC experiments upgrades~\cite{CERN-LHCC-2017-023}, and considered by the CLIC collaboration. The CLIC ECAL barrel, on which the current muon collider detector design is based, is composed of 64M sensors sampling 40 layers. Future developments should implement a precise timing measurement in these sensors ($<$100 ps) in order to make them usable at a muon collider. Although the high granularity is a clear advantage, the associated number of electronic read-out channels is a non-trivial technological problem. Moreover the cost for such a system exceeds the cost of other solutions. 

\subsubsection{Scintillator-based sampling}
Plastic scintillators can be used for a high-granularity detector. Small tiles or strips of scintillating material can be produced with a typical size of 1-5~cm$^{2}$ for e.m. applications and about 10~cm$^{2}$ for hadronic applications. The typical thickness of the active layer is 0.3-0.5~cm, which makes it possible to design detectors with a longitudinal segmentation of several tens of layers. Each calorimeter cell can be read out via a silicon-based photo-detector mounted directly on the scintillating element~\cite{Sefkow:2018rhp}. Passive absorbers, such as steel, can be intertwined with the sensitive layers.
The high granularity that can be achieved allows to implement compensation in hadronic showers in the reconstruction software using energy density techniques.

\subsubsection{Micro-pattern gaseous detector-based sampling}
Calorimeters with gaseous detectors as active element can reach a higher granularity with respect to more traditional scintillator-based calorimeters. Furthermore, a 1x1~cm$^2$ pad is economically affordable with respect to 3x3~cm$^2$ scintillator tiles. The CALICE collaboration has been studying the performance of digital~\cite{Adams:2016por} and semi-digital~\cite{Baulieu:2015pfa} hadronic calorimeters. Besides having a high granularity, gaseous detectors have the advantage to be radiation hard, leading to simple calibration procedures. RPCs have been chosen historically because of their low cost to instrument large area, because of their intrinsic digital nature and because of the rather low rate expected at future electron-positron colliders, which were the main target of these designs. Recently, Micro-Pattern Gaseous Detectors (MPGDs) have been proposed since they will likely outperform Resistive-Plate Chambers (RPCs) for this task because of their higher rate capability, their operation with environmental-friendly gases, their good energy resolution (of about 20\%), high detector stability and low pad multiplicity. In the last decade resistive Micromegas were developed and tested for calorimetry at ILC~\cite{Chefdeville:2021ava} and for tracking in high-rate environments~\cite{Alviggi:2020hoy} at LHC, while new resistive detectors like the $\mu$RWELL~\cite{Bencivenni:2014exa} and RP-WELL~\cite{Rubin:2013jna} were developed, the latter being actively pursued for SDHCAL calorimeter of CepC detector~\cite{ShakedRenous:2020xeu}. 
It is expected that a good time resolution will play an important role in helping to match tracking detector and calorimeter energy deposits. Single-layer RPCs can reach time-resolutions of few hundred ps, but rely on gases with high global warming potential that will be phased out in the near future. Replacement gases are currently under investigation. Micro-Pattern Gaseous Detectors typically have a time resolution of the order of few ns, with R\&D ongoing to bring this to sub-ns.

\subsection{Other promising calorimeter technologies}

Calorimeters are generally divided in two categories, homogeneous and sampling. The best compromise between the two technologies is sought in order to optimize experimental requirements and minimize the drawbacks associated with the limitations of standard solutions. The most recent technological developments allow this rigid distinction to be abandoned in favour of novel architectures: the Crilin calorimeter is a semi-homogeneous calorimeter based on Lead Fluoride (\pbfd) crystals readout by surface mounted UV extended Silicon Photomultipliers (SiPMs).

\subsubsection{Crilin: a CRystal calorImeter with Longitudinal INformation}

The Crilin calorimeter can be segmented longitudinally as a function of the energy of the particles and of the background level, thanks to its modular design which enables a high degree of reconfigurability. The Crilin R\&D proposal embeds a modular architecture based on stackable submodules composed of matrices of  crystals, in which each crystal is individually readout by 2 series of 2 UV-extended surface mount SiPMs.  Crystal dimensions are $10\times10\times40$~mm$^3$ and the surface area of each SiPM is 4$\times$4~mm$^2$, so as to closely match the crystal surface.

\begin{figure}[h!]
    \centering
    \includegraphics[width=0.9\textwidth]{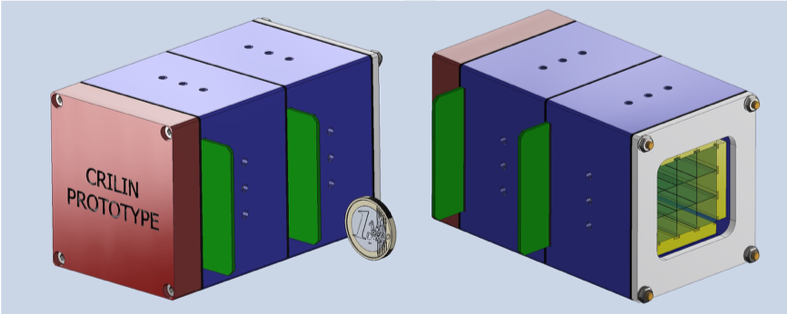}
    \caption{CAD model of  Crilin Prototype.}
    \label{fig:x1}
\end{figure}

In the current design, the prototype consists of two submodules, each composed of a 3-by-3 crystals matrix. The submodules are arranged in a series and assembled together by screws, resulting in a compact and small calorimeter, shown in Figure~\ref{fig:x1}.

Each crystal matrix is housed in a light-tight case which also embeds the front-end electronic boards and the cooling system. The on-detector electronics and SiPMs must be cooled during operation, so as to improve and stabilise the performance of SiPMs against irradiation. The Crilin design is capable of removing the heat load due to the increased photosensor currents after exposure to the expected $2-5 \cdot 10^{13}$~1-MeV-neq~cm$^{-2}$/year fluence, along with the power dissipated by the amplification circuitry. The total heat load was estimated as 350~mW per channel. The Crilin cooling system, which is based on conduction and forced convection of nitrogen, will provide the optimum operating temperature for the electronics and SiPMs at around $0^{\circ}$. Gas fluxing will also prevent any condensation on SiPM or crystal surfaces.
\FloatBarrier

\section{Muon systems}
\label{sec:muon}
A compact detector can be obtained using an iron yoke to concentrate the magnetic flux return from the solenoid, instrumented with several layers of gaseous detectors.

\begin{figure}[!ht]
    \center
    \includegraphics[width=0.48\textwidth]{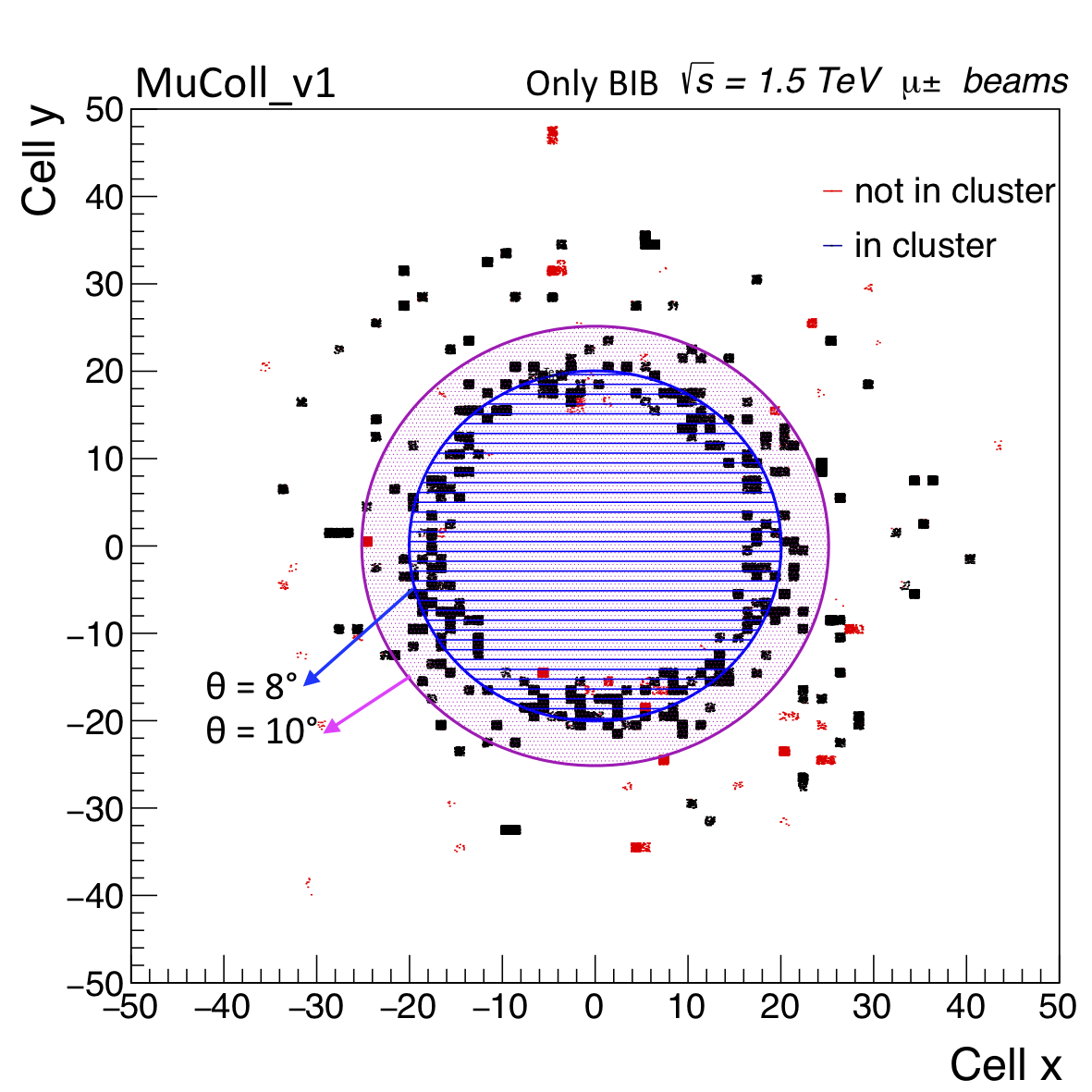}
    \caption{Left: BIB muon hit spatial distribution in the first layer of the muon endcap. The detector hits not associated to a cluster are shown by the red markers. The blue circle corresponds to region $\theta<8^{\circ}$, while the purple to $\theta<10^{\circ}$.}
    \label{fig:BIB2fig}
\end{figure}

The BIB in the muon system is mainly composed of high energy neutrons and photons.
\begin{figure}[!h]
    \center
    \includegraphics[width=0.5\textwidth]{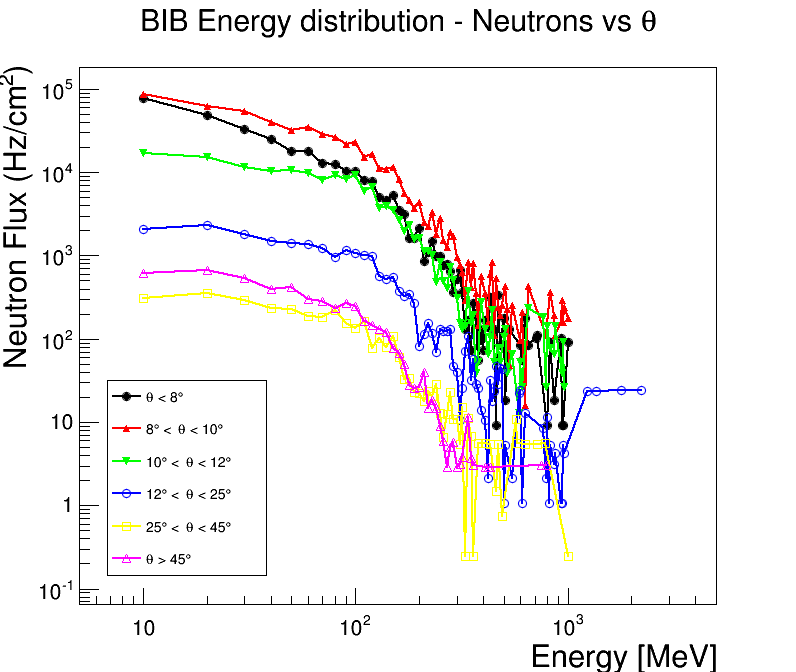}\includegraphics[width=0.5\textwidth]{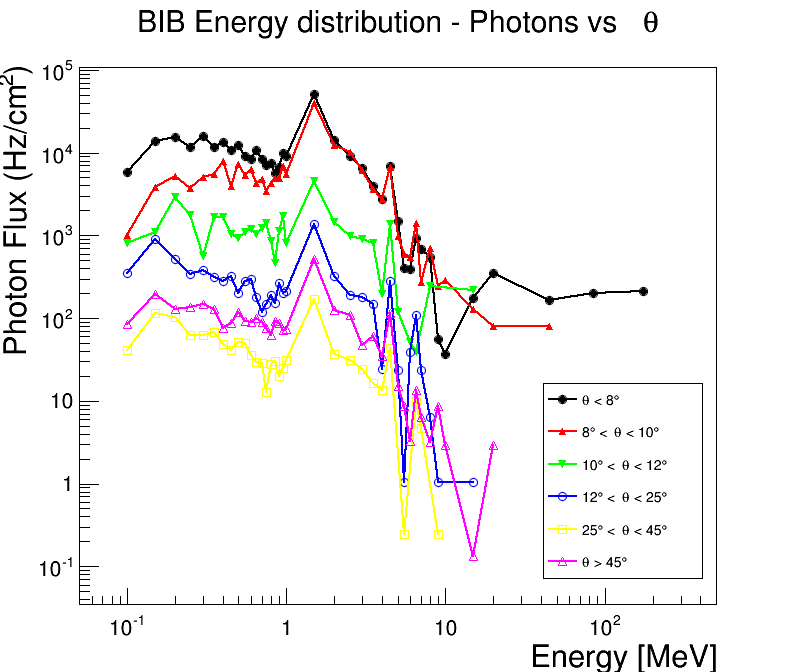}
    \caption{Energy distribution of neutrons (left) and photons (right) from BIB at a 1.5~TeV muon collider. Different colors represent different geometrical regions of the muon system in the endcap.}
    \label{fig:BIB_15TeV_fluxes}
\end{figure}
The energy spectra of both the components are shown in Fig.~\ref{fig:BIB_15TeV_fluxes}: neutron energy ranges from \SI{10}{MeV} up to \SI{2.5}{GeV}, with the majority of the flux in the region $E < 100$~MeV; photons instead go from \SI{100}{keV} to \SI{200}{MeV}, with the majority of the flux in the region $E < $\SI{10}{MeV}.
The colors represent different geometrical regions of the detector endcap, based on the angular coordinate $\theta$ (or $r$): as expected, the fluxes are higher in the inner part of the endcap, at lower $\theta$ and closer to the beam line, and then lower in the outer regions. The same subdivision has been used then to evaluate the expected hit rate in the system.
BIB hits in the muon system are, indeed, concentrated around the beam axis in the endcaps in a region small with respect to the whole layer region of 500$\times$500 cells, as shown in Figure~\ref{fig:BIB2fig}.

Sensitivity studies pointed out that some technologies, such as RPCs, are already at the limit of their current rate capability. These results, together with preliminary requirements on the spatial ($\approx$\SI{100}{\micro \meter}) and time resolution (below \SI{1}{ns}), suggest the need for gaseous detectors R\&D. Classical well-known MPGDs, such as GEMs or Micromegas, are characterized by an excellent spatial resolution, but do not match the demanding request on the timing resolution. R\&D on new generation MPGDs are still at the initial phase, but are obtaining promising results for a future implementation. A possible muon system design can be a heterogeneous detector, composed of layers of different technologies to optimize the timing and tracking performance. Moreover, an excellent spatial resolution would give the possibility to use the standalone muon objects to seed the global muon track reconstruction.

\subsection{Detectors with high spatial resolution}
\label{detec_space}
Classical gaseous detectors, Multi-Wire Proportional Chambers (MWPCs), reach spatial resolutions of the order of few mm (dominated by the mechanical limitation in the wire spacing). Resolutions of \SI{50}{\micro m} to \SI{100}{\micro m} are obtained measuring precisely the drift time (Drift Tubes, DT) or by patterning the cathode combined with precise charge measurement (Cathode Strip Chambers, CSCs), however all wire-based detectors have intrinsic rate limitations due to the slow evacuation of ions. This limitation has been overcome in Micro-pattern Gaseous Detectors, where the electrodes are created using photo-lithographic techniques, which allows the reduction of the electrode spacing of at least one order of magnitude, resulting in fast ion evacuation combined with high spatial resolution.

\subsubsection{Gas Electron Multiplier}
A Gas Electron Multiplier (GEM) consists of a thin polymer foil (often \SI{50}{\micro m} polyimide), cladded on both sides with a thin layer of copper (\SI{5}{\micro m}), chemically perforated with a high density of holes, typically 100/mm$^2$~\cite{Sauli:1997qp}. Applying a potential difference between the top and bottom electrodes, a high electric field is formed in the holes where electron multiplication can take place. Several GEM-foils can be stacked on top of each other leading to detectors with high gain ($>10^5$) and low discharge probability ($<10^{-10}$). GEM detectors are used in different collider experiments~\cite{Ketzer:2001dt}, mainly for tracking and triggering purposes. Spatial resolution down to \SI{50}{\micro \meter} is possible, with a time resolution that depends on the used gas mixture used: 7--10~ns in Ar:CO2~\cite{CMSMuon:2019pzw} down to 3.5~ns when CF$_4$ is used in the gas mixture~\cite{Alfonsi:2004gh}. Rate capabilities up to 100 kHz/cm$^2$ have been assessed.

\subsubsection{Micromegas}
Micromegas are parallel-plate chambers where the amplification takes  place in a thin gap, separated from the conversion region by a fine metallic mesh~\cite{Giomataris:1995fq}. Micromegas are used in collider experiments~\cite{Thers:2001qs} and a spark-protected evolution with resistive strips will be used mainly for tracking in the upgraded forward muon system of the ATLAS experiment~\cite{ATLAS-TDR-020}. A spatial resolution of \SI{80}{\micro \meter}, a time resolution of 7-\SI{10}{ns}, and a rate capability up to \SI{100}{kHz/cm^2} have been achieved.

\subsubsection{Micro-resistive Wells}
The Micro-resistive Well ($\mu$-RWELL) is a single-amplification stage resistive MPGD~\cite{Bencivenni:2014exa}. Such technology is reliable, since the presence of the resistive layer assures a very low discharge rate quenching the spark amplitude. It can achieve a rate capability up to 10~MHz/cm$^2$ with a detection efficiency of the order of 97-98\%~\cite{Bencivenni:2019wxr}. Typical spatial resolution is $<$\SI{60}{\micro \meter}, time resolution measured with CF$_4$ gas mixture was measured to be below \SI{6}{ns}, time resolution in Ar:CO$_2$ mixtures is expected to be similar to the triple-GEM detectors (7-10\,ns).

\subsection{Detectors with sub-ns timing resolution}
\label{detec_timing}
\subsubsection{Multigap RPC}

MRPC~\cite{CerronZeballos:1995iy} have acquired solidity and importance in the High Energy Physics domain where both high efficiency and  good time resolution are demanding. A time resolution of about \SI{60}{ps} and $95\%$ efficiency has been obtained~\cite{ALICE:2008ngc,Llope:2005yw} for a detector composed of 10 gas gaps $250~\mu$m size arranged in a double stack design using floating soda-lime glass (bulk resistivity $ 5 \times 10^{12}\, \Omega\, cm$).  A recirculating gas mixture (C$_2$H$_2$F$_4$:SF$_6$ 97:3) allows operation in avalanche mode with electric field of approximately $100$~kV/cm. The technology was operated~\cite{Friese:2006dj} in high particle fluxes (though for small-size detector units) in a series of tests for the endcap upgrade of the STAR ToF and for the mini-CBM  experiment at the GSI/SIS8 Synchrotron.  Data with thinner standard float glass show rate capability of some~kHz/cm$^2$ at time resolutions of 70--80~ps.

The rate capability is limited by the current flowing through the resistive plates and hence a step forward to increase this value, at a constant front end electronics threshold, would be to use low resistivity glass or to decrease their thickness. Lower resistivity glass ($ 10^{10}\,\Omega\,$cm) have been used in tests beam~\cite{Wang:2019jjz} and showed rate capability of 35~kHz/cm$^2$ with efficiency above $90\%$ and time resolutions below \SI{80}{ps}. A still better behavior can be seen by lowering the glass resistivity by one order of magnitude, currently R\&D is concentrated on establishing \SI{20}{ps} time resolution at rates of 100~kHz/cm$^2$.

Although the MRPC have shown excellent timing resolution over large area and R\&D for larger rate capability and even better timing is underway, this detector technology can not be considered for future collider experiments with the current gas mixture which has a high Global Warming Potential (GWP). The use of freons will be gradually phased out by 2030. While for standard (High Pressure Laminate, HPL) RPCs encouraging results have been obtained to replace the freons with alternative gases as Hydrofluoroolefine (HFO-1234ze), MRPC performances have yet to be proven with those new gas mixtures. Moreover MRPC performance relies on a non-negligible fraction of SF$_6$ which has even higher GWP with respect to the freons. R\&D should start urgently in order to propose these detectors for future use.

\subsubsection{Micropattern Gaseous Detectors with improved time resolution}
The time resolution of a classical MPGD is dominated by the fluctuations on the position on the first ionization cluster in the drift gap.
The contribution to the time resolution of the drift velocity ($v_d$) is then given by $\sigma_t = (\lambda v_d)^{-1}$ where $\lambda$ is the average number of primary clusters generated by an ionizing particle inside the gas per unit length~\cite{DeOliveira:2015bda}. Therefore a better time resolution is expected with a faster mixture. However, even with a fast gas mixture, as for example Ar:CO$_2$:CF$_4$, classical MPGDs usually cannot reach time resolution better than few ns.

A possible approach aimed at improving the time resolution is the one followed by the PicoSec Collaboration~\cite{Bortfeldt:2019ecw}: the proposed detector is a standard Micromegas (MM) with a reduced drift gap (down to \SI{200}{\micro \meter}) in order to minimize the possibility of primary ionization. Particles pass instead through a Cherenkov radiator placed on top of the MM, where they produce Cherenkov photons, which are then converted by a photocathode and enter in the drift region, removing the fluctuations on the position of the first ionization cluster.
Preliminary results of the first prototypes of few~cm$^2$ proved the reliability of the principle with a time resolution of \SI{25}{ps} measured with a gas mixture of Ne/C$_2$H$_6$/CF$_4$. Further studies of this new technology will be focused on: proper stability of the detectors, choice of the materials and geometry, radiation hardness and gas mixture (avoid use of CF$_4$ and flammable gases).

An alternative approach is represented by the Fast Timing Micropattern (FTM) gas detector~\cite{DeOliveira:2015bda}. Here the drift gap is segmented in $N$ thinner fully resistive drift+amplification stages, which are in competition between each other when the detector is fully efficient (sum of drift gaps $\geq$ \SI{1.5}{mm}). The fastest stage is the one that determines the timing of the signal, thus reducing the time resolution of the detector of a factor $N$ with respect to a standard MPGD. The first prototype, made of just two stages, obtained a time resolution of \SI{2}{ns} with pion beam~\cite{Abbaneo:2017tat}. The current R\&D is focused on improving the quality of the resistive layers used for the amplification stage and establishing the detection principle with multiple layers on a small scale prototype~\cite{Colaleo:2019tmy}, and assessing the technology on larger scale prototypes. The main limitations come from the quality of the detector elements with resistive layers, which is a technology currently developed, and the single-stage reachable gain; interesting results have been recently obtained with a Ne/iC$_4$H$_{10}$\,95/5 gas mixture~\cite{Pellecchia:2021sip}.

\subsection{Technology comparison}
In order to understand the response of the muon detectors to the BIB particles, the detector sensitivities have been studied with a standalone Geant4 simulation.

\begin{figure}[!ht]
    \center
    \includegraphics[width=0.5\textwidth]{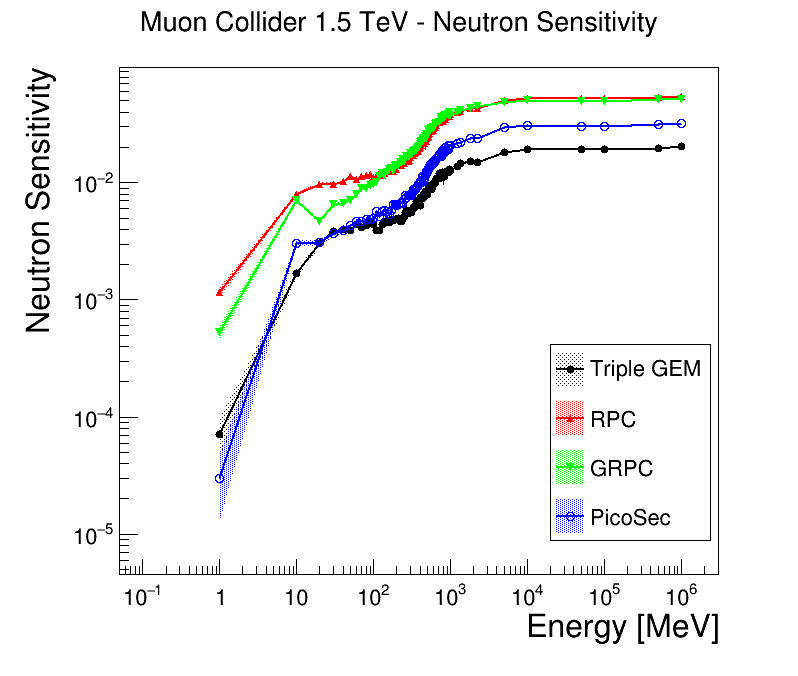}\includegraphics[width=0.5\textwidth]{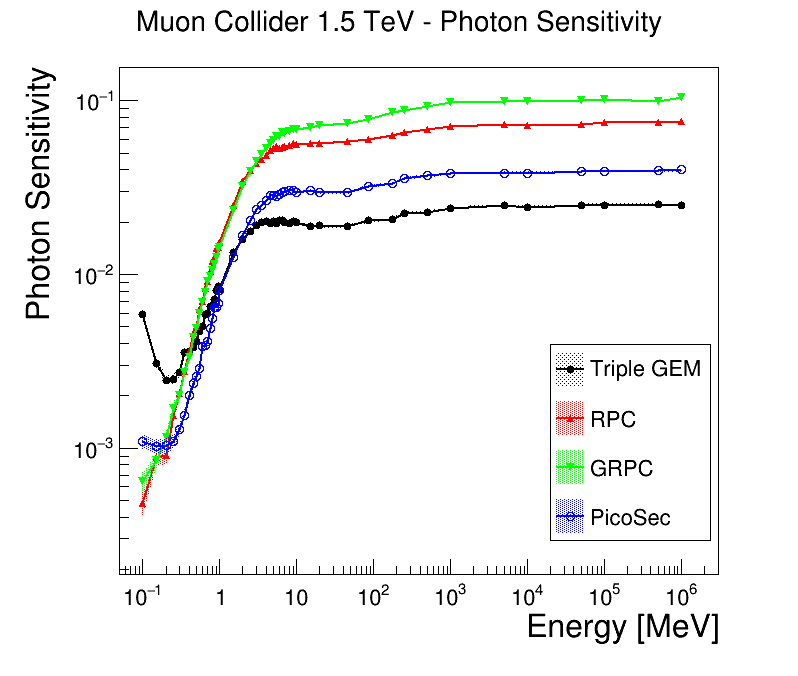}
    \vspace{-0.50cm}
    \caption{Simulated sensitivity to neutrons (left) and photons (right). Different colors represent different gaseous detector technologies considered.}
    \label{fig:BIB_15TeV_sensitivities}
\end{figure}

Sensitivity may be defined as the probability for a BIB particle to generate a visible signal in the detector: in the simulation, it is computed as the ratio $s = N/M$, where $N$ is the number of events in which at least one charged particle reaches a sensitive gas gap, while $M$ is the number of incident particles. Figure~\ref{fig:BIB_15TeV_sensitivities} shows the results of the simulation for \SI{1.5}{TeV} muon collider BIB neutrons (left) and photons (right) for four different gaseous detector technologies considered and described in Sections~\ref{detec_space}-\ref{detec_timing}.

The energy range over which the simulation was performed covers completely the BIB range and extends it, in view of possible modifications of the spectra at higher centre of mass energies. The difference between the technologies is mainly due to a different material composition of the detectors: in general we can observe that Micro-Pattern Gaseous Detectors, i.e. Triple-GEM (Gaseous Electron Multiplier) and PicoSec, result in having a lower sensitivity both to neutrons and photons with respect to RPC (both Glass RPC and standard HPL RPC).

\begin{figure}[!ht]
    \center
    \includegraphics[width=0.5\textwidth]{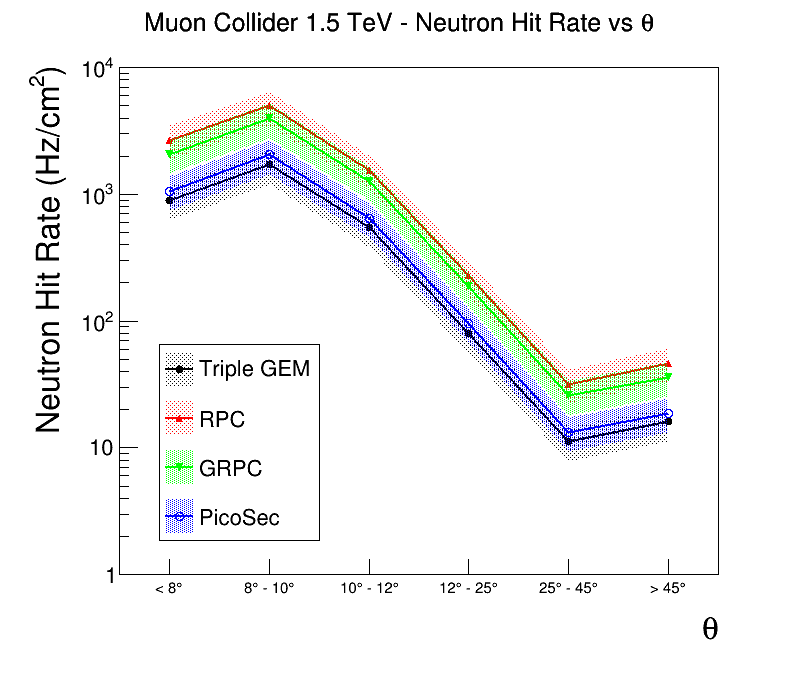}\includegraphics[width=0.5\textwidth]{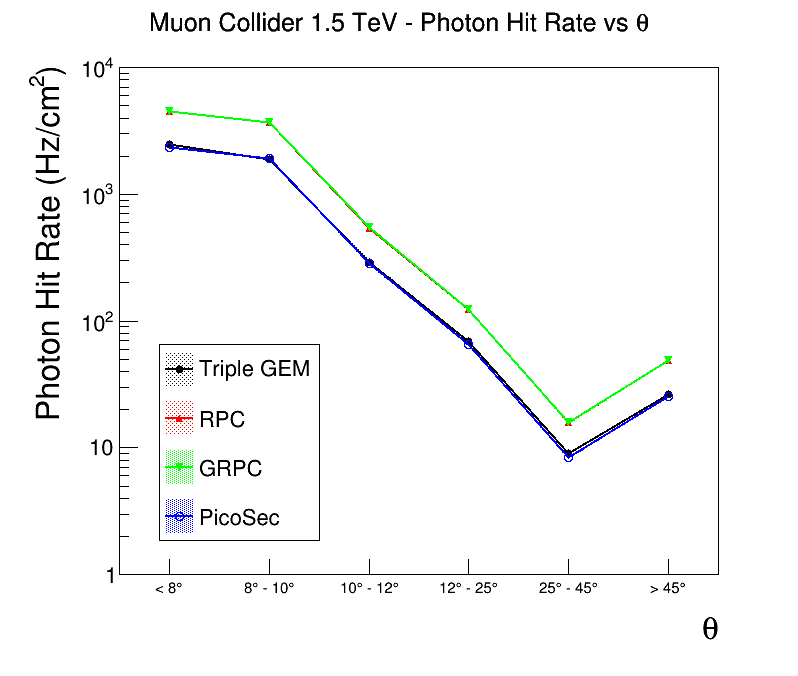}
    \vspace{-0.50cm}
    \caption{Estimated hit rate from neutrons (left) and photons (right) at a \SI{1.5}{TeV} muon collider. Different colors represent different gaseous detector technologies considered.}
    \label{fig:BIB_15TeV_hitrates}
\end{figure}

The hit rate (R) is then obtained from the flux ($\Phi$) by $R = s \times \Phi$ for each energy value and particle type. The total estimated rate is shown in Fig.~\ref{fig:BIB_15TeV_hitrates} for neutrons (left) and photons (right) as a function of the angular coordinate $\theta$ for the different detector technologies considered.

As expected, the estimated hit rate is higher in the inner endcap and decreases with $\theta$, i.e. when moving to the outer region of the endcap.

\FloatBarrier

\section{Trigger and Data Acquisition}
\label{sec:tdaq}

Experiments at the high energy muon collider are expected to operate at instantaneous luminosity levels of $10^{34}-10^{35}$ cm$^{-2}$ s$^{-1}$. With a single bunch collider operation scheme, beam crossing frequency is defined by the beam energy and the size of the collider ring. Collisions  are expected to happen at the maximum rate of 100 kHz, corresponding to the minimum time between crossings of 10 $\mu$s.

The Trigger and Data Acquisition (TDAQ) systems of the future muon collider experiments will be required to perform partial or full reconstruction of every collision event in order to identify and store events of interest to the physics programme. Given that the earliest a muon collider can be realized is 20 years from now, it is way too early to define the TDAQ strategy. Future advancements in the data transmission and processing technologies can substantially alter the vision of what a TDAQ system at the muon collider would look like. However, an initial estimate of the data rates and processing needs does help to outline possible options and strategies, in particular when put in the context of today’s technologies. 

Trigger and DAQ strategy taken by different collider experiments varies a lot and depends on the luminosity and complexity of their collision events. Experiments such as ATLAS and CMS, utilize hardware triggers~\cite{ATLAS:2016wtr,CMS:2016ngn} that rely on a subset of the detector subsystems for initial filtering of the events. This is followed by a High Level Trigger (HLT) farm where further processing and filtering takes place using more complete event information. LHCb experiment operating at lower luminosity and with smaller event size recently opted for a so-called “triggerless” or “streaming” approach~\cite{LHCbCollaboration:2014vzo,Colombo:2018upq} which eliminates need for a hardware trigger and where all collision data is streamed at 40 MHz directly to an HLT farm for event reconstruction. Similarly, electron-positron collider experiments~\cite{Behnke:2013lya,CLICdp:2018vnx} typically adopt a triggerless readout scheme due to the relative cleanliness of events when compared to hadron colliders. A streaming approach offers a number of advantages: availability of the full event data typically translates into a better trigger decision, it is easier to support and upgrade software triggers, simplified design of the detector front-end, etc. However, the presence of large BIB at the muon collider may be prohibitive for a full triggerless TDAQ scheme. In the following, an initial estimate of the data rates is provided to show that from the data rates/volumes consideration a streaming DAQ implementation is feasible. We also provide early estimates of the event processing time and compare it to that anticipated at the HL-LHC experiments.

The amount of data acquired by the muon collider experiments is expected to be dominated by the tracker and calorimeters. For the silicon tracker, we estimate event size and data rates by acquiring an average number of hits per event from simulation and multiplying it by the 100 kHz event rate. The average hit multiplicity as a function of the tracker layer can be found in Figure~\ref{fig:HitMultTracker}. We assume that each hit consists of 32 bits to encode charge, position, and time information and that zero-suppression is applied in the detector front-end. Hits are integrated in the time period of 1 ns following the bunch crossing, which allows to preserve good efficiency for hits from particles originating in the hard scattering but rejects a significant fraction of the BIB. No filtering based on the hit directionality information is applied to avoid possible biases in the online selection. An additional safety factor of 2 is embedded in the calculation. With this, we estimate the tracker event size to be 40 Megabytes (MB) and the data rate from the tracker to be 30 Terabits per second (Tb/s). It should be noted that the numbers are dominated by the BIB hits. A similar approach is applied for calculating event data rates originating in the calorimeter. Here the ECAL dominates with approximately 90 million channels and average occupancy of about 10$^{-3}$ hits per mm$^{2}$. A minimum energy threshold of 0.2~MeV is required in order for the hit to be read out and hits are assumed to be 20 bits wide. The HCAL contribution is small, less than 10\% of the ECAL.  After applying a safety factor of 2, calorimeter event size is estimated to be 40 MB and similar to that for the tracker. The full data rate corresponding to the sum of the tracker and calorimeter rates is therefore about 60 Tb/s, which is a factor of few larger than HLT input of LHCb experiment in Run-3~\cite{LHCbCollaboration:2014vzo} and comparable to the HLT input of CMS experiment in HL-LHC~\cite{Collaboration:2759072}. Therefore, from the data volumes point of view, a streaming operation at 100 kHz appears to be feasible. It should be emphasized that the rates are directly proportional to the bunch crossing frequency and can be much larger with a smaller collider ring or a multi-bunch operation scheme, in which case the strategy may have to be re-evaluated.

Another important parameter to consider is the HLT output rate to offline storage. Here again different approaches can be taken. One approach is to eliminate most of the events using filtering done in the HLT, but preserve full raw event information for the ones that pass the filter.  HL-LHC experiments assume HLT output bandwidth of approximately 60 GB/s. This would translate into 750 Hz of full event content sent to the storage. For comparison, single Higgs production at 10 TeV is expected to have a much lower rate of <0.1 Hz and WW production via the vector boson fusion will be at 1 Hz level. Storing full event content allows for later reprocessing of the data in order to improve the performance or reconstruct novel signatures. However, this approach would require about 4 PB of storage per day of running. It will produce a total dataset similar in size to that of ATLAS and CMS in HL-LHC, but dominated by BIB hits that are not interesting from the physics point of view. An alternative approach based on reducing event content by filtering out hits and clusters clearly unassociated with the hard scattering may be more prudent. Here one may choose to preserve the output rate of 750 kHz and reduce the total dataset size. Alternatively, one can aim for the same total dataset size but increase the rate of events, for example if on average 99\% of BIB hits in each event can be filtered out, every produced event at 100~kHz can be sent to the storage. 

In addition to data rates, it is also important to take into account time needed to reconstruct each event at the HLT.  Long processing times lead to unmanageable large farm requirements, which in turn is difficult to procure, maintain, and operate. HL-LHC experiments project processing times of about 1 second per event with tails extending as far as a minute. At the muon collider, offline event reconstruction is currently dominated by charged particle tracking and takes up to few minutes per event using Kalman filter based tracking algorithms from the ACTS package, which is significantly slower than a good HLT target of few seconds. To estimate the amount of time needed at the HLT, a preliminary reconstruction of charged tracks was attempted in the outer layers where BIB density is less severe. To have a first clue on the complexity of the problem we took hits from the outer six barrels and apply an algorithm based on a three-dimensional Hough Transform in the ($\phi$, $\eta$, $p_T$) parameter space and assuming all tracks coming from the origin as a good approximation at this stage. Preliminary analytical calculations suggest that, even with a three-dimensional array finely subdivided into 240 million cells, with currently estimated BIB rates, tens of purely combinatorial track candidates with $p_T$ > 2.5 GeV/c would be found, precluding any possibility to use that information as an effective event filter. However, if an independent filter based on timing (or pointing, etc ...) is applied to lower the BIB hit rate by a factor four, requiring a single candidate track with $p_T > 2.5$ GeV would give us the needed event rejection factor of about 50. In this case, 44 million cells in the Hough Transform array would have to be filled for each beam crossing. Doing this in few seconds requires a powerful CPU, but is not out of reach. Future CPUs and accelerators (GPU, FPGA, etc), as well as algorithm improvements can further improve timing. The bottom line is that this is a crucial area of development and careful attention needs to be paid to this area in order to keep the processing time under control.

\begin{figure}[!ht]
\center
\includegraphics[width=0.45\textwidth]{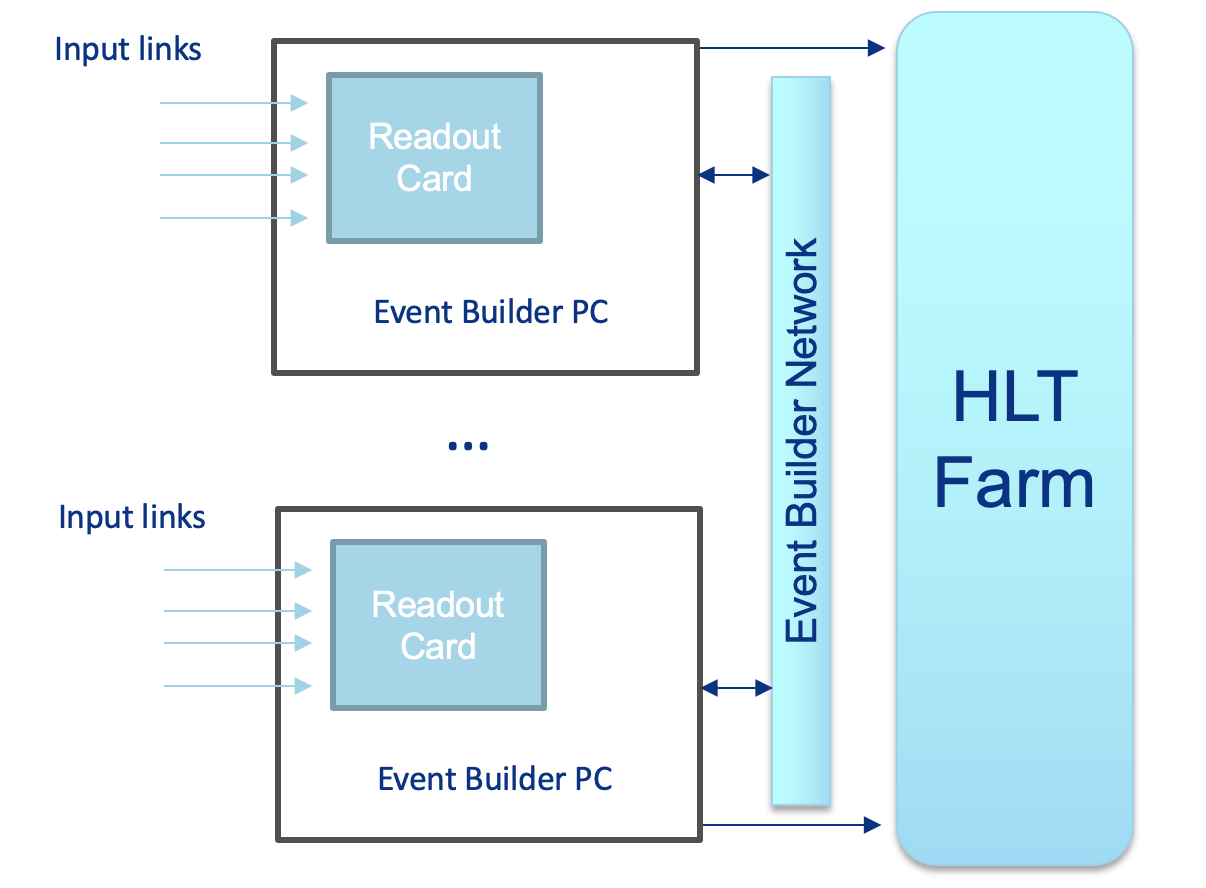}
\includegraphics[width=0.45\textwidth]{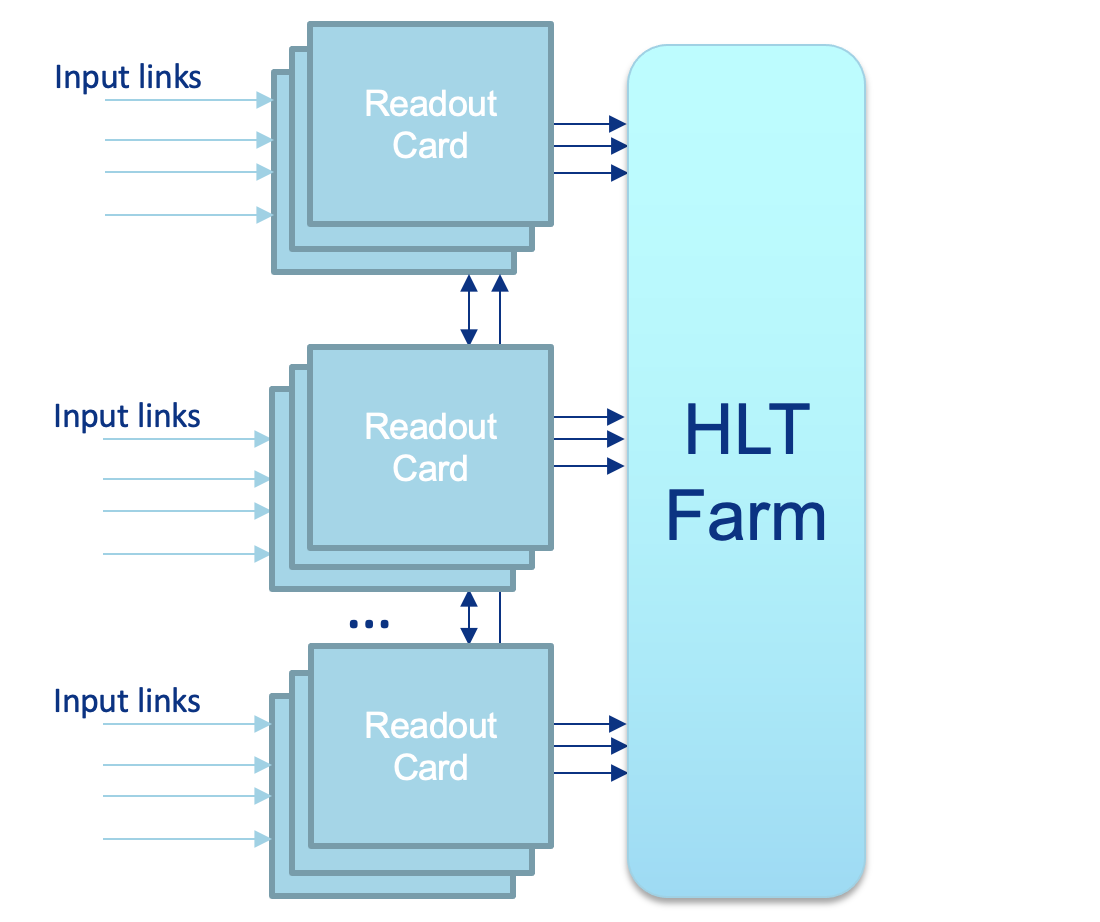}
\caption{Schematic illustration of possible ways to structure the TDAQ system. The one on the left shows an LHCb-like approach with a software Event Builder. The one on the right uses hardware boards to structure event data and pass them to the HLT farm.}
\label{fig:TDAQArch}
\end{figure}

Finally, it may be useful to sketch how a potential TDAQ system could look and estimate its size. The number of input links is an important parameter for this purpose. lpGBT developed for HL-LHC are expected to provide bandwidth of up to 10~Gb/s per link. It is reasonable to assume that links with at least twice higher speeds (20~Gb/s) will be available on the timescale of 20 years. We thus assume that data from multiple detector modules can be combined into a single 20~Gb/s link and estimate that close to 10,000 of such links will be needed to bring the data from the detector to the electronics area. The back-end system will consist of few hundred readout boards used to receive and format the data. A full-mesh hardware or a software Event Builder network with aggregate bandwidth of about 60~Tb/s will be required, for which both custom and industry provided solutions can be considered. Note that this number can be much smaller if a more aggressive filtering scheme than the one described above is utilized in the front-end. For example, preliminary studies indicate that filtering tracker hits based on their pointing information can yield up to a factor of eight reduction in the data rates. The required output bandwidth to storage will depend on the chosen filtering strategy at the HLT, but should not exceed 100~Gb/s even in the most aggressive scenarios. A schematic illustration with a couple of different architectures can be found in Fig.~\ref{fig:TDAQArch}, with LHCb-like approach shown on the left and a full-mesh FPGA hardware event builder approach on the right. 

To summarize, despite large BIB, preliminary estimates based on simulation indicate that a streaming DAQ architecture can provide an attractive solution for the future muon collider experiments. With improvements to the tracking speed, such a solution can likely be realized with technologies available today. Future  advancements (e.g. higher speed optical links, fast processors, etc) are likely to result in a smaller and/or more performing DAQ system. Work should be invested in improving  HLT reconstruction algorithms and exploiting GPU/FPGA/ASIC acceleration schemes with the aim to bring per-event processing time down to a few second level.

\section{Conclusions}
\label{sec:conclusions}
The 2019 update of the European Strategy for Particle Physics and the 2021 Snowmass process identified muon colliders as a highly promising path to reaching very high center-of-mass energies in leptonic collisions. These machines can combine excellent new physics discovery potential with high precision capabilities. For the purpose of event detection and reconstruction, the challenge that separates $\mu^{+}\mu^{-}$ with the $e^{+}e^{-}$ counterparts is the beam induced background. Because muons are unstable, they decay in flight, producing electrons that further interact with the accelerator and detector components. This creates very large multiplicities of mostly soft secondary particles, some of which end up in the detector. 
The hits produced by the secondary particles in the detectors lead to a large  challenge for the particle detection and reconstruction. In this paper we presented a preliminary design and specifications for a muon collider detector, described the radiation levels it will be exposed to, and outlined possible ways to mitigate effects of the BIB. We showed that the radiation levels are similar to those at the HL-LHC experiments. We also demonstrated that the BIB imposes stringent requirements on the granularity, resolution, and timing properties of the muon collider detectors, and presented a number of emerging detector technologies that have a potential to address the challenge. Finally, we argue that R\&D efforts to further develop these technologies are needed to get them to the maturity levels necessary for the detector construction.


\FloatBarrier
\bibliographystyle{report}
\bibliography{MC_Detector_technologies.bib}

\providecommand{\href}[2]{#2}\begingroup\raggedright\begin{thebibliography}{100}

\bibitem{sm21-imcc-accelerator-summary}
{International Muon Collider Collaboration}, {\em A short summary of the muon
  collider facility prospects\/},  To Appear  .

\bibitem{Delahaye:2019omf}
J.~P. Delahaye et al., {\em {Muon Colliders}\/},
  \href{http://arxiv.org/abs/1901.06150}{{\tt arXiv:1901.06150
  [physics.acc-ph]}}.

\bibitem{Long:2020wfp}
K.~Long et al., {\em {Muon Colliders: Opening New Horizons for Particle
  Physics}\/},  \href{http://arxiv.org/abs/2007.15684}{{\tt arXiv:2007.15684
  [physics.acc-ph]}}.

\bibitem{tikhonin1968effects}
F.~Tikhonin and G.~Budker, {\em On the effects with muon colliding beams\/},
  JINR Report P2-4120 (1968)  33–39.

\bibitem{budker1982accelerators}
G.~Budker, {\em Accelerators and colliding beams\/},  Proc. 7th International
  Conference on High-Energy Accelerators Vol. 1 (1970)  .

\bibitem{Delahaye:2013jla}
J.-P. Delahaye et al., {\em {Enabling Intensity and Energy Frontier Science
  with a Muon Accelerator Facility in the U.S.: A White Paper Submitted to the
  2013 U.S. Community Summer Study of the Division of Particles and Fields of
  the American Physical Society}\/},
  \href{http://arxiv.org/abs/1308.0494}{{\tt arXiv:1308.0494
  [physics.acc-ph]}}.

\bibitem{sm21-imcc-physics-summary}
{International Muon Collider Collaboration}, {\em Muon Collider Physics
  Summary\/},  To Appear  .

\bibitem{sm21-imcc-physics-3TeV}
{International Muon Collider Collaboration}, {\em The physics case of a 3~TeV
  muon collider stage\/},  To Appear  .

\bibitem{Buttazzo:2018qqp}
D.~Buttazzo, D.~Redigolo, F.~Sala, and A.~Tesi, {\em {Fusing Vectors into
  Scalars at High Energy Lepton Colliders}\/},
  \href{http://dx.doi.org/10.1007/JHEP11(2018)144}{JHEP {\bf 11} (2018)  144},
  \href{http://arxiv.org/abs/1807.04743}{{\tt arXiv:1807.04743 [hep-ph]}}.

\bibitem{Franceschini:2021dxn}
R.~Franceschini, {\em {High(est) energy lepton colliders}\/},
  \href{http://dx.doi.org/10.1142/S0217751X21420161}{Int. J. Mod. Phys. A {\bf
  36} (2021) no.~22, 2142016}.

\bibitem{Capdevilla:2021fmj}
R.~Capdevilla, F.~Meloni, R.~Simoniello, and J.~Zurita, {\em {Hunting wino and
  higgsino dark matter at the muon collider with disappearing tracks}\/},
  \href{http://dx.doi.org/10.1007/JHEP06(2021)133}{JHEP {\bf 06} (2021)  133},
  \href{http://arxiv.org/abs/2102.11292}{{\tt arXiv:2102.11292 [hep-ph]}}.

\bibitem{AlAli:2021let}
H.~Al~Ali et al., {\em {The Muon Smasher's Guide}\/},
  \href{http://arxiv.org/abs/2103.14043}{{\tt arXiv:2103.14043 [hep-ph]}}.

\bibitem{Buttazzo:2020uzc}
D.~Buttazzo, R.~Franceschini, and A.~Wulzer, {\em {Two Paths Towards Precision
  at a Very High Energy Lepton Collider}\/},
  \href{http://arxiv.org/abs/2012.11555}{{\tt arXiv:2012.11555 [hep-ph]}}.

\bibitem{Han:2020pif}
T.~Han, D.~Liu, I.~Low, and X.~Wang, {\em {Electroweak Couplings of the Higgs
  Boson at a Multi-TeV Muon Collider}\/},
  \href{http://dx.doi.org/10.1103/PhysRevD.103.013002}{Phys. Rev. D {\bf 103}
  (2021)  013002}, \href{http://arxiv.org/abs/2008.12204}{{\tt arXiv:2008.12204
  [hep-ph]}}.

\bibitem{Capdevilla:2020qel}
R.~Capdevilla, D.~Curtin, Y.~Kahn, and G.~Krnjaic, {\em {A Guaranteed Discovery
  at Future Muon Colliders}\/},  \href{http://arxiv.org/abs/2006.16277}{{\tt
  arXiv:2006.16277 [hep-ph]}}.

\bibitem{Ruhdorfer:2019utl}
M.~Ruhdorfer, E.~Salvioni, and A.~Weiler, {\em {A Global View of the Off-Shell
  Higgs Portal}\/},
  \href{http://dx.doi.org/10.21468/SciPostPhys.8.2.027}{SciPost Phys. {\bf 8}
  (2020)  027}, \href{http://arxiv.org/abs/1910.04170}{{\tt arXiv:1910.04170
  [hep-ph]}}.

\bibitem{Chiesa:2020awd}
M.~Chiesa et al., {\em {Measuring the quartic Higgs self-coupling at a
  multi-TeV muon collider}\/},
  \href{http://dx.doi.org/10.1007/JHEP09(2020)098}{JHEP {\bf 09} (2020)  098},
  \href{http://arxiv.org/abs/2003.13628}{{\tt arXiv:2003.13628 [hep-ph]}}.

\bibitem{Costantini:2020stv}
A.~Costantini et al., {\em {Vector boson fusion at multi-TeV muon
  colliders}\/},  \href{http://dx.doi.org/10.1007/JHEP09(2020)080}{JHEP {\bf
  09} (2020)  080}, \href{http://arxiv.org/abs/2005.10289}{{\tt
  arXiv:2005.10289 [hep-ph]}}.

\bibitem{Capdevilla:2021rwo}
R.~Capdevilla, D.~Curtin, Y.~Kahn, and G.~Krnjaic, {\em {A No-Lose Theorem for
  Discovering the New Physics of $(g-2)_\mu$ at Muon Colliders}\/},
  \href{http://arxiv.org/abs/2101.10334}{{\tt arXiv:2101.10334 [hep-ph]}}.

\bibitem{Bartosik:2020xwr}
N.~Bartosik et al., {\em {Detector and Physics Performance at a Muon
  Collider}\/},  \href{http://dx.doi.org/10.1088/1748-0221/15/05/P05001}{JINST
  {\bf 15} (2020) no.~05, P05001}, \href{http://arxiv.org/abs/2001.04431}{{\tt
  arXiv:2001.04431 [hep-ex]}}.

\bibitem{Yin:2020afe}
W.~Yin and M.~Yamaguchi, {\em {Muon $g-2$ at multi-TeV muon collider}\/},
  \href{http://arxiv.org/abs/2012.03928}{{\tt arXiv:2012.03928 [hep-ph]}}.

\bibitem{Kalinowski:2020rmb}
J.~Kalinowski, T.~Robens, D.~Sokolowska, and A.~F. Zarnecki, {\em {IDM
  Benchmarks for the LHC and Future Colliders}\/},
  \href{http://dx.doi.org/10.3390/sym13060991}{Symmetry {\bf 13} (2021) no.~6,
  991}, \href{http://arxiv.org/abs/2012.14818}{{\tt arXiv:2012.14818
  [hep-ph]}}.

\bibitem{Liu:2021jyc}
W.~Liu and K.-P. Xie, {\em {Probing electroweak phase transition with multi-TeV
  muon colliders and gravitational waves}\/},
  \href{http://dx.doi.org/10.1007/JHEP04(2021)015}{JHEP {\bf 04} (2021)  015},
  \href{http://arxiv.org/abs/2101.10469}{{\tt arXiv:2101.10469 [hep-ph]}}.

\bibitem{Han:2021udl}
T.~Han, S.~Li, S.~Su, W.~Su, and Y.~Wu, {\em {Heavy Higgs Bosons in 2HDM at a
  Muon Collider}\/},  \href{http://arxiv.org/abs/2102.08386}{{\tt
  arXiv:2102.08386 [hep-ph]}}.

\bibitem{Bottaro:2021srh}
S.~Bottaro, A.~Strumia, and N.~Vignaroli, {\em {Minimal Dark Matter bound
  states at future colliders}\/},
  \href{http://dx.doi.org/10.1007/JHEP06(2021)143}{JHEP {\bf 06} (2021)  143},
  \href{http://arxiv.org/abs/2103.12766}{{\tt arXiv:2103.12766 [hep-ph]}}.

\bibitem{Li:2021lnz}
T.~Li, M.~A. Schmidt, C.-Y. Yao, and M.~Yuan, {\em {Charged lepton flavor
  violation in light of the muon magnetic moment anomaly and colliders}\/},
  \href{http://dx.doi.org/10.1140/epjc/s10052-021-09569-9}{Eur. Phys. J. C {\bf
  81} (2021) no.~09, 811}, \href{http://arxiv.org/abs/2104.04494}{{\tt
  arXiv:2104.04494 [hep-ph]}}.

\bibitem{Asadi:2021gah}
P.~Asadi, R.~Capdevilla, C.~Cesarotti, and S.~Homiller, {\em {Searching for
  leptoquarks at future muon colliders}\/},
  \href{http://dx.doi.org/10.1007/JHEP10(2021)182}{JHEP {\bf 10} (2021)  182},
  \href{http://arxiv.org/abs/2104.05720}{{\tt arXiv:2104.05720 [hep-ph]}}.

\bibitem{Sahin:2021xzt}
M.~Sahin and A.~Caliskan, {\em {Excited muon production at muon colliders via
  contact interaction}\/},  \href{http://arxiv.org/abs/2105.01964}{{\tt
  arXiv:2105.01964 [hep-ph]}}.

\bibitem{Chen:2021rid}
J.~Chen, C.-T. Lu, and Y.~Wu, {\em {Measuring Higgs boson self-couplings with 2
  \textrightarrow{} 3 VBS processes}\/},
  \href{http://dx.doi.org/10.1007/JHEP10(2021)099}{JHEP {\bf 10} (2021)  099},
  \href{http://arxiv.org/abs/2105.11500}{{\tt arXiv:2105.11500 [hep-ph]}}.

\bibitem{Haghighat:2021djz}
G.~Haghighat and M.~Mohammadi~Najafabadi, {\em {Search for
  lepton-flavor-violating ALPs at a future muon collider and utilization of
  polarization-induced effects}\/},
  \href{http://arxiv.org/abs/2106.00505}{{\tt arXiv:2106.00505 [hep-ph]}}.

\bibitem{Bottaro:2021snn}
S.~Bottaro et al., {\em {Closing the window on WIMP Dark Matter}\/},
  \href{http://arxiv.org/abs/2107.09688}{{\tt arXiv:2107.09688 [hep-ph]}}.

\bibitem{Sen:2021fha}
C.~Sen, P.~Bandyopadhyay, S.~Dutta, and A.~KT, {\em {Displaced Higgs production
  in Type-III Seesaw at the LHC/FCC, MATHUSLA and Muon collider}\/},
  \href{http://arxiv.org/abs/2107.12442}{{\tt arXiv:2107.12442 [hep-ph]}}.

\bibitem{Han:2021lnp}
T.~Han, W.~Kilian, N.~Kreher, Y.~Ma, J.~Reuter, T.~Striegl, and K.~Xie, {\em
  {Precision test of the muon-Higgs coupling at a high-energy muon
  collider}\/},  \href{http://dx.doi.org/10.1007/JHEP12(2021)162}{JHEP {\bf 12}
  (2021)  162}, \href{http://arxiv.org/abs/2108.05362}{{\tt arXiv:2108.05362
  [hep-ph]}}.

\bibitem{Bandyopadhyay:2021pld}
P.~Bandyopadhyay, A.~Karan, and R.~Mandal, {\em {Distinguishing signatures of
  scalar leptoquarks at hadron and muon colliders}\/},
  \href{http://arxiv.org/abs/2108.06506}{{\tt arXiv:2108.06506 [hep-ph]}}.

\bibitem{Dermisek:2021mhi}
R.~Dermisek, K.~Hermanek, and N.~McGinnis, {\em {Di-Higgs and tri-Higgs boson
  signals of muon g-2 at a muon collider}\/},
  \href{http://dx.doi.org/10.1103/PhysRevD.104.L091301}{Phys. Rev. D {\bf 104}
  (2021) no.~9, L091301}, \href{http://arxiv.org/abs/2108.10950}{{\tt
  arXiv:2108.10950 [hep-ph]}}.

\bibitem{Qian:2021ihf}
S.~Qian, C.~Li, Q.~Li, F.~Meng, J.~Xiao, T.~Yang, M.~Lu, and Z.~You, {\em
  {Searching for heavy leptoquarks at a muon collider}\/},
  \href{http://arxiv.org/abs/2109.01265}{{\tt arXiv:2109.01265 [hep-ph]}}.

\bibitem{Chiesa:2021qpr}
M.~Chiesa, B.~Mele, and F.~Piccinini, {\em {Multi Higgs production via photon
  fusion at future multi-TeV muon colliders}\/},
  \href{http://arxiv.org/abs/2109.10109}{{\tt arXiv:2109.10109 [hep-ph]}}.

\bibitem{Liu:2021akf}
W.~Liu, K.-P. Xie, and Z.~Yi, {\em {Testing leptogenesis at the LHC and future
  muon colliders: a $Z'$ scenario}\/},
  \href{http://arxiv.org/abs/2109.15087}{{\tt arXiv:2109.15087 [hep-ph]}}.

\bibitem{Buttazzo:2021lzi}
D.~Buttazzo and P.~Paradisi, {\em {Probing the muon $g-2$ anomaly with the
  Higgs boson at a muon collider}\/},
  \href{http://dx.doi.org/10.1103/PhysRevD.104.075021}{Phys. Rev. D {\bf 104}
  (2021) no.~7, 075021}.

\bibitem{DiLuzio:2018jwd}
L.~Di~Luzio, R.~Gr\"ober, and G.~Panico, {\em {Probing new electroweak states
  via precision measurements at the LHC and future colliders}\/},
  \href{http://dx.doi.org/10.1007/JHEP01(2019)011}{JHEP {\bf 01} (2019)  011},
  \href{http://arxiv.org/abs/1810.10993}{{\tt arXiv:1810.10993 [hep-ph]}}.

\bibitem{Han:2020uak}
T.~Han, Z.~Liu, L.-T. Wang, and X.~Wang, {\em {WIMPs at High Energy Muon
  Colliders}\/},  \href{http://dx.doi.org/10.1103/PhysRevD.103.075004}{Phys.
  Rev. D {\bf 103} (2021) no.~7, 075004},
  \href{http://arxiv.org/abs/2009.11287}{{\tt arXiv:2009.11287 [hep-ph]}}.

\bibitem{Chen:2021pqi}
J.~Chen, T.~Li, C.-T. Lu, Y.~Wu, and C.-Y. Yao, {\em {The measurement of Higgs
  self-couplings through $2\rightarrow 3$ VBS in future muon colliders}\/},
  \href{http://arxiv.org/abs/2112.12507}{{\tt arXiv:2112.12507 [hep-ph]}}.

\bibitem{Capdevilla:2021kcf}
R.~Capdevilla, D.~Curtin, Y.~Kahn, and G.~Krnjaic, {\em {Systematically Testing
  Singlet Models for $(g-2)_\mu$}\/},
  \href{http://arxiv.org/abs/2112.08377}{{\tt arXiv:2112.08377 [hep-ph]}}.

\bibitem{Cesarotti:2022ttv}
C.~Cesarotti, S.~Homiller, R.~K. Mishra, and M.~Reece, {\em {Probing New Gauge
  Forces with a High-Energy Muon Beam Dump}\/},
  \href{http://arxiv.org/abs/2202.12302}{{\tt arXiv:2202.12302 [hep-ph]}}.

\bibitem{sm21-imcc-accelerator}
{International Muon Collider Collaboration}, {\em A Muon Collider Facility for
  Physics Discovery\/},  To Appear  .

\bibitem{MICE}
{MICE collaboration}, {\em {Demonstration of cooling by the Muon Ionization
  Cooling Experiment}\/},
  \href{http://dx.doi.org/https://doi.org/10.1038/s41586-020-1958-9}{Nature
  {\bf 578} (2020)  53–59}.

\bibitem{Alesini:2019tlf}
D.~Alesini et al., {\em {Positron driven muon source for a muon collider}\/},
  \href{http://arxiv.org/abs/1905.05747}{{\tt arXiv:1905.05747
  [physics.acc-ph]}}.

\bibitem{sm21-imcc-performance}
{International Muon Collider Collaboration}, {\em Simulated detector
  performance at the Muon Collider\/},  To Appear  .

\bibitem{Mokhov:2011zzd}
N.~V. Mokhov and S.~I. Striganov, {\em {Detector Background at Muon
  Colliders}\/},  \href{http://dx.doi.org/10.1016/j.phpro.2012.03.761}{Phys.
  Procedia {\bf 37} (2012)  2015--2022},
  \href{http://arxiv.org/abs/1204.6721}{{\tt arXiv:1204.6721
  [physics.ins-det]}}.

\bibitem{Mokhov:2014hza}
N.~V. Mokhov, S.~I. Striganov, and I.~S. Tropin,
  \href{http://dx.doi.org/10.18429/JACoW-IPAC2014-TUPRO029}{{\em {Reducing
  Backgrounds in the Higgs Factory Muon Collider Detector}\/}, } in {\em {5th
  International Particle Accelerator Conference}}, pp.~1081--1083.
\newblock 6, 2014.
\newblock \href{http://arxiv.org/abs/1409.1939}{{\tt arXiv:1409.1939
  [physics.ins-det]}}.

\bibitem{Bartosik:2019dzq}
N.~Bartosik et al., {\em {Preliminary Report on the Study of Beam-Induced
  Background Effects at a Muon Collider}\/},
  \href{http://arxiv.org/abs/1905.03725}{{\tt arXiv:1905.03725 [hep-ex]}}.

\bibitem{Lucchesi:2020dku}
D.~Lucchesi et al., {\em {Detector Performances Studies at Muon Collider}\/},
  \href{http://dx.doi.org/10.22323/1.364.0118}{PoS {\bf EPS-HEP2019} (2020)
  118}.

\bibitem{Collamati:2021sbv}
F.~Collamati et al., {\em {Advanced assessment of beam-induced background at a
  muon collider}\/},
  \href{http://dx.doi.org/10.1088/1748-0221/16/11/P11009}{JINST {\bf 16} (2021)
  no.~11, P11009}, \href{http://arxiv.org/abs/2105.09116}{{\tt arXiv:2105.09116
  [physics.acc-ph]}}.

\bibitem{Mokhov:2017klc}
N.~V. Mokhov and C.~C. James, {\em {The MARS Code System User\textquoteright{}s
  Guide Version 15(2016)}\/},  {FERMILAB-FN-1058-APC}, 2017.

\bibitem{Ahdida:2022gjl}
C.~Ahdida et al., {\em {New Capabilities of the FLUKA Multi-Purpose Code}\/},
  \href{http://dx.doi.org/10.3389/fphy.2021.788253}{Front. in Phys. {\bf 9}
  (2022)  788253}.

\bibitem{Battistoni:2015epi}
G.~Battistoni et al., {\em {Overview of the FLUKA code}\/},
  \href{http://dx.doi.org/10.1016/j.anucene.2014.11.007}{Annals Nucl. Energy
  {\bf 82} (2015)  10--18}.

\bibitem{CLICdp:2017vju}
{CLICdp Collaboration}, {\em {CLICdet: The post-CDR CLIC detector model}\/},
  {CLICdp-Note-2017-001}, 2017.
\newblock \url{https://cds.cern.ch/record/2254048}.

\bibitem{CLICdp:2018vnx}
D.~Arominski et al., {\em {A detector for CLIC: main parameters and
  performance}\/},  \href{http://arxiv.org/abs/1812.07337}{{\tt
  arXiv:1812.07337 [physics.ins-det]}}.

\bibitem{ILDConceptGroup:2020sfq}
{ILD Concept Group Collaboration}, H.~Abramowicz et al., {\em {International
  Large Detector: Interim Design Report}\/},
  \href{http://arxiv.org/abs/2003.01116}{{\tt arXiv:2003.01116
  [physics.ins-det]}}.

\bibitem{ILC:2007vrf}
{ILC Collaboration}, G.~Aarons et al., {\em {ILC Reference Design Report Volume
  4 - Detectors}\/},  \href{http://arxiv.org/abs/0712.2356}{{\tt
  arXiv:0712.2356 [physics.ins-det]}}.

\bibitem{Lipton:2022njd}
R.~J. Lipton, {\em {A Double Sided LGAD-Based Detector Providing Timing,
  Position, and Track Angle Information}\/},  {FERMILAB-FN-1102-E}, 2022.

\bibitem{Lipton:2019drv}
R.~Lipton and J.~Theiman, {\em {Fast timing with induced current detectors}\/},
   \href{http://dx.doi.org/10.1016/j.nima.2019.162423}{Nucl. Instrum. Meth. A
  {\bf 945} (2019)  162423}.

\bibitem{Cardella:2019ksc}
R.~Cardella et al., {\em {MALTA: an asynchronous readout CMOS monolithic pixel
  detector for the ATLAS High-Luminosity upgrade}\/},
  \href{http://dx.doi.org/10.1088/1748-0221/14/06/C06019}{JINST {\bf 14} (2019)
  no.~06, C06019}.

\bibitem{Giacomini:2019kqz}
G.~Giacomini, W.~Chen, G.~D'Amen, and A.~Tricoli, {\em {Fabrication and
  performance of AC-coupled LGADs}\/},
  \href{http://dx.doi.org/10.1088/1748-0221/14/09/p09004}{JINST {\bf 14} (2019)
  no.~09, P09004}, \href{http://arxiv.org/abs/1906.11542}{{\tt arXiv:1906.11542
  [physics.ins-det]}}.

\bibitem{Tornago:2020otn}
M.~Tornago et al., {\em {Resistive AC-Coupled Silicon Detectors: principles of
  operation and first results from a combined analysis of beam test and laser
  data}\/},  \href{http://dx.doi.org/10.1016/j.nima.2021.165319}{Nucl. Instrum.
  Meth. A {\bf 1003} (2021)  165319},
  \href{http://arxiv.org/abs/2007.09528}{{\tt arXiv:2007.09528
  [physics.ins-det]}}.

\bibitem{Lipton:2015vca}
R.~Lipton et al., {\em {Three Dimensional Integrated Circuits Bonded to
  Sensors}\/},  \href{http://dx.doi.org/10.22323/1.227.0045}{PoS {\bf
  Vertex2014} (2015)  045}.

\bibitem{OConnor:1999isd}
P.~O'Connor and G.~De~Geronimo,
  \href{http://dx.doi.org/10.1109/NSSMIC.1999.842454}{{\em {Prospects for
  charge sensitive amplifiers in scaled CMOS}\/}, } in {\em {IEEE Nuclear
  Science Symposium and Medical Imaging Conference}}, IEEE
  Nucl.Sci.Symp.Conf.Rec., pp.~88--93.
\newblock 2000.

\bibitem{RLDDesign}
{R. Lipton}, {\em {Muon Collider Detector Design}\/},  2020.
\newblock \url{https://indico.cern.ch/event/969833/}.

\bibitem{lpGBT}
P.~Moreira, {\em {The LpGBT Project, Status and Overview}\/},  2016.
\newblock \url{https://indico.cern.ch/event/468486/contributions/1144369/}.

\bibitem{7805240}
D.~A.~B. Miller, {\em Attojoule Optoelectronics for Low-Energy Information
  Processing and Communications\/},
  \href{http://dx.doi.org/10.1109/JLT.2017.2647779}{Journal of Lightwave
  Technology {\bf 35} (2017) no.~3, 346--396}.

\bibitem{Lee:2017shn}
S.~Lee et al., {\em {Hadron detection with a dual-readout fiber
  calorimeter}\/},  \href{http://dx.doi.org/10.1016/j.nima.2017.05.025}{Nucl.
  Instrum. Meth. A {\bf 866} (2017)  76--90},
  \href{http://arxiv.org/abs/1703.09120}{{\tt arXiv:1703.09120
  [physics.ins-det]}}.

\bibitem{dr20}
M.~Antonello et al., {\em {Dual-readout calorimetry, an integrated
  high-resolution solution for energy measurements at future
  electron\textendash{}positron colliders}\/},
  \href{http://dx.doi.org/10.1016/j.nima.2019.04.017}{Nucl. Instrum. Meth. A
  {\bf 958} (2020)  162063}.

\bibitem{b}
{RD52 Collaboration}, R.~Ferrari, {\em {Dual-readout calorimetry: recent
  results from RD52 and plans for experiments at future $e^{+}e^{-}$
  colliders}\/},  \href{http://dx.doi.org/10.1088/1748-0221/13/02/C02050}{JINST
  {\bf 13} (2018) no.~02, C02050}.

\bibitem{Thomson:2009rp}
M.~A. Thomson, {\em {Particle Flow Calorimetry and the PandoraPFA
  Algorithm}\/},  \href{http://dx.doi.org/10.1016/j.nima.2009.09.009}{Nucl.
  Instrum. Meth. A {\bf 611} (2009)  25--40},
  \href{http://arxiv.org/abs/0907.3577}{{\tt arXiv:0907.3577
  [physics.ins-det]}}.

\bibitem{Wigmans:2007es}
R.~Wigmans, {\em {The DREAM project: Results and plans}\/},
  \href{http://dx.doi.org/10.1016/j.nima.2006.10.211}{Nucl. Instrum. Meth. A
  {\bf 572} (2007)  215--217}.

\bibitem{FCC:2018evy}
A.~Abada et al., {\em {FCC-ee: The Lepton Collider}: {Future Circular Collider
  Conceptual Design Report Volume 2}\/},
  \href{http://dx.doi.org/10.1140/epjst/e2019-900045-4}{Eur. Phys. J. ST {\bf
  228} (2019) no.~2, 261--623}.

\bibitem{CEPCStudyGroup:2018rmc}
{CEPC Study Group}, {\em {CEPC Conceptual Design Report: Volume 1 -
  Accelerator}\/},  \href{http://arxiv.org/abs/1809.00285}{{\tt
  arXiv:1809.00285 [physics.acc-ph]}}.

\bibitem{Lee:2017xss}
S.~Lee, M.~Livan, and R.~Wigmans, {\em {Dual-Readout Calorimetry}\/},
  \href{http://dx.doi.org/10.1103/RevModPhys.90.025002}{Rev. Mod. Phys. {\bf
  90} (2018) no.~2, 025002}, \href{http://arxiv.org/abs/1712.05494}{{\tt
  arXiv:1712.05494 [physics.ins-det]}}.

\bibitem{Lee:2017oye}
S.~Lee, M.~Livan, and R.~Wigmans, {\em {On the limits of the hadronic energy
  resolution of calorimeters}\/},
  \href{http://dx.doi.org/10.1016/j.nima.2017.10.087}{Nucl. Instrum. Meth. A
  {\bf 882} (2018)  148--157}, \href{http://arxiv.org/abs/1710.10535}{{\tt
  arXiv:1710.10535 [physics.ins-det]}}.

\bibitem{Lucchini:2020bac}
M.~T. Lucchini et al., {\em {New perspectives on segmented crystal calorimeters
  for future colliders}\/},
  \href{http://dx.doi.org/10.1088/1748-0221/15/11/P11005}{JINST {\bf 15} (2020)
  no.~11, P11005}, \href{http://arxiv.org/abs/2008.00338}{{\tt arXiv:2008.00338
  [physics.ins-det]}}.

\bibitem{Gaudio:2011zzb}
{DREAM Collaboration}, G.~Gaudio, {\em {New results from the DREAM project}\/},
   \href{http://dx.doi.org/10.1016/j.nima.2010.06.348}{Nucl. Instrum. Meth. A
  {\bf 628} (2011)  339--342}.

\bibitem{Brient:2001fow}
J.-C. Brient and H.~Videau, {\em {The Calorimetry at the future e$^{+}$e$^{-}$
  linear collider}\/},  eConf {\bf C010630} (2001)  E3047,
  \href{http://arxiv.org/abs/hep-ex/0202004}{{\tt arXiv:hep-ex/0202004}}.

\bibitem{CERN-LHCC-2017-023}
{CMS Collaboration}, {\em {The Phase-2 Upgrade of the CMS Endcap
  Calorimeter}\/},  {CMS-TDR-019}, 2017.
\newblock \url{https://cds.cern.ch/record/2293646}.

\bibitem{Sefkow:2018rhp}
{CALICE Collaboration}, F.~Sefkow and F.~Simon, {\em {A highly granular
  SiPM-on-tile calorimeter prototype}\/},
  \href{http://dx.doi.org/10.1088/1742-6596/1162/1/012012}{J. Phys. Conf. Ser.
  {\bf 1162} (2019) no.~1, 012012}, \href{http://arxiv.org/abs/1808.09281}{{\tt
  arXiv:1808.09281 [physics.ins-det]}}.

\bibitem{Adams:2016por}
C.~Adams et al., {\em {Design, construction and commissioning of the Digital
  Hadron Calorimeter\textemdash{}DHCAL}\/},
  \href{http://dx.doi.org/10.1088/1748-0221/11/07/P07007}{JINST {\bf 11} (2016)
  no.~07, P07007}, \href{http://arxiv.org/abs/1603.01653}{{\tt arXiv:1603.01653
  [physics.ins-det]}}.

\bibitem{Baulieu:2015pfa}
G.~Baulieu et al., {\em {Construction and commissioning of a technological
  prototype of a high-granularity semi-digital hadronic calorimeter}\/},
  \href{http://dx.doi.org/10.1088/1748-0221/10/10/P10039}{JINST {\bf 10} (2015)
  no.~10, P10039}, \href{http://arxiv.org/abs/1506.05316}{{\tt arXiv:1506.05316
  [physics.ins-det]}}.

\bibitem{Chefdeville:2021ava}
M.~Chefdeville et al., {\em {Development of Micromegas detectors with resistive
  anode pads}\/},  \href{http://dx.doi.org/10.1016/j.nima.2021.165268}{Nucl.
  Instrum. Meth. A {\bf 1003} (2021)  165268}.

\bibitem{Alviggi:2020hoy}
M.~Alviggi et al., {\em {Pixelated resistive bulk micromegas for tracking
  systems in high rate environment}\/},
  \href{http://dx.doi.org/10.1088/1748-0221/15/06/C06035}{JINST {\bf 15} (2020)
  no.~06, C06035}.

\bibitem{Bencivenni:2014exa}
G.~Bencivenni, R.~De~Oliveira, G.~Morello, and M.~P. Lener, {\em {The
  micro-Resistive WELL detector: a compact spark-protected single
  amplification-stage MPGD}\/},
  \href{http://dx.doi.org/10.1088/1748-0221/10/02/P02008}{JINST {\bf 10} (2015)
  no.~02, P02008}, \href{http://arxiv.org/abs/1411.2466}{{\tt arXiv:1411.2466
  [physics.ins-det]}}.

\bibitem{Rubin:2013jna}
A.~Rubin et al., {\em {First studies with the Resistive-Plate WELL gaseous
  multiplier}\/},  \href{http://dx.doi.org/10.1088/1748-0221/8/11/P11004}{JINST
  {\bf 8} (2013)  P11004}, \href{http://arxiv.org/abs/1308.6152}{{\tt
  arXiv:1308.6152 [physics.ins-det]}}.

\bibitem{ShakedRenous:2020xeu}
D.~Shaked~Renous et al., {\em {Towards MPGD-based (S)DHCAL}\/},
  \href{http://dx.doi.org/10.1088/1742-6596/1498/1/012040}{J. Phys. Conf. Ser.
  {\bf 1498} (2020)  012040}.

\bibitem{Sauli:1997qp}
F.~Sauli, {\em {GEM: A new concept for electron amplification in gas
  detectors}\/},  \href{http://dx.doi.org/10.1016/S0168-9002(96)01172-2}{Nucl.
  Instrum. Meth. A {\bf 386} (1997)  531--534}.

\bibitem{Ketzer:2001dt}
B.~Ketzer et al., {\em {Triple GEM tracking detectors for COMPASS}\/},
  \href{http://dx.doi.org/10.1109/TNS.2002.803891}{IEEE Trans. Nucl. Sci. {\bf
  49} (2002)  2403--2410}.

\bibitem{CMSMuon:2019pzw}
M.~Abbas et al., {\em {Performance of prototype GE1 \ensuremath{/} 1 chambers
  for the CMS muon spectrometer upgrade}\/},
  \href{http://dx.doi.org/10.1016/j.nima.2020.164104}{Nucl. Instrum. Meth. A
  {\bf 972} (2020)  164104}, \href{http://arxiv.org/abs/1903.02186}{{\tt
  arXiv:1903.02186 [physics.ins-det]}}.

\bibitem{Alfonsi:2004gh}
M.~Alfonsi et al., {\em {High-rate particle triggering with triple-GEM
  detector}\/},  \href{http://dx.doi.org/10.1016/j.nima.2003.10.035}{Nucl.
  Instrum. Meth. A {\bf 518} (2004)  106--112}.

\bibitem{Giomataris:1995fq}
Y.~Giomataris, P.~Rebourgeard, J.~P. Robert, and G.~Charpak, {\em {MICROMEGAS:
  A High granularity position sensitive gaseous detector for high particle flux
  environments}\/},
  \href{http://dx.doi.org/10.1016/0168-9002(96)00175-1}{Nucl. Instrum. Meth. A
  {\bf 376} (1996)  29--35}.

\bibitem{Thers:2001qs}
D.~Thers et al., {\em {Micromegas as a large microstrip detector for the
  COMPASS experiment}\/},
  \href{http://dx.doi.org/10.1016/S0168-9002(01)00769-0}{Nucl. Instrum. Meth. A
  {\bf 469} (2001)  133--146}.

\bibitem{ATLAS-TDR-020}
{ATLAS Collaboration}, {\em {New Small Wheel Technical Design Report}\/},
  {ATLAS-TDR-020}, Jun, 2013.
\newblock \url{http://cds.cern.ch/record/1552862}.

\bibitem{Bencivenni:2019wxr}
G.~Bencivenni et al., {\em {The $\mu$-RWELL layouts for high particle rate}\/},
   \href{http://dx.doi.org/10.1088/1748-0221/14/05/P05014}{JINST {\bf 14}
  (2019) no.~05, P05014}, \href{http://arxiv.org/abs/1903.11017}{{\tt
  arXiv:1903.11017 [physics.ins-det]}}.

\bibitem{CerronZeballos:1995iy}
C.~Zeballos et al., {\em {A New type of resistive plate chamber: The Multigap
  RPC}\/},  \href{http://dx.doi.org/10.1016/0168-9002(96)00158-1}{Nucl.
  Instrum. Meth. A {\bf 374} (1996)  132--136}.

\bibitem{ALICE:2008ngc}
{ALICE Collaboration}, {\em {The ALICE experiment at the CERN LHC}\/},
  \href{http://dx.doi.org/10.1088/1748-0221/3/08/S08002}{JINST {\bf 3} (2008)
  S08002}.

\bibitem{Llope:2005yw}
W.~J. Llope, {\em {The large-area time-of-flight upgrade for STAR}\/},
  \href{http://dx.doi.org/10.1016/j.nimb.2005.07.089}{Nucl. Instrum. Meth. B
  {\bf 241} (2005)  306--310}.

\bibitem{Friese:2006dj}
V.~Friese, {\em {The CBM experiment at GSI/FAIR}\/},
  \href{http://dx.doi.org/10.1016/j.nuclphysa.2006.06.018}{Nucl. Phys. A {\bf
  774} (2006)  377--386}.

\bibitem{Wang:2019jjz}
Y.~Wang et al., {\em {Development and Production of High Rate MRPC for CBM
  TOF}\/},  \href{http://dx.doi.org/10.7566/JPSCP.26.024006}{JPS Conf. Proc.
  {\bf 26} (2019)  024006}.

\bibitem{DeOliveira:2015bda}
R.~De~Oliveira, M.~Maggi, and A.~Sharma, {\em {A novel fast timing micropattern
  gaseous detector: FTM}\/},  \href{http://arxiv.org/abs/1503.05330}{{\tt
  arXiv:1503.05330 [physics.ins-det]}}.

\bibitem{Bortfeldt:2019ecw}
J.~Bortfeldt et al., {\em {Precise Charged Particle Timing with the PICOSEC
  Detector}\/},  \href{http://dx.doi.org/10.1063/1.5091210}{AIP Conf. Proc.
  {\bf 2075} (2019) no.~1, 080009}, \href{http://arxiv.org/abs/1901.03355}{{\tt
  arXiv:1901.03355 [physics.ins-det]}}.

\bibitem{Abbaneo:2017tat}
D.~Abbaneo et al., {\em {R\&D on a new type of micropattern gaseous detector:
  The Fast Timing Micropattern detector}\/},
  \href{http://dx.doi.org/10.1016/j.nima.2016.05.067}{Nucl. Instrum. Meth. A
  {\bf 845} (2017)  313--317}.

\bibitem{Colaleo:2019tmy}
A.~Colaleo et al., {\em {Diamond-Like Carbon for the Fast Timing MPGD}\/},
  \href{http://dx.doi.org/10.1088/1742-6596/1498/1/012015}{J. Phys. Conf. Ser.
  {\bf 1498} (2020) no.~1, 012015}, \href{http://arxiv.org/abs/1907.13559}{{\tt
  arXiv:1907.13559 [physics.ins-det]}}.

\bibitem{Pellecchia:2021sip}
A.~Pellecchia and P.~Verwilligen, {\em {Performance of a fast timing
  micro-pattern gaseous detector for future collider experiments}\/},
  \href{http://arxiv.org/abs/2107.09439}{{\tt arXiv:2107.09439
  [physics.ins-det]}}.

\bibitem{ATLAS:2016wtr}
{ATLAS Collaboration}, {\em {Performance of the ATLAS Trigger System in
  2015}\/},  \href{http://dx.doi.org/10.1140/epjc/s10052-017-4852-3}{Eur. Phys.
  J. C {\bf 77} (2017) no.~5, 317}, \href{http://arxiv.org/abs/1611.09661}{{\tt
  arXiv:1611.09661 [hep-ex]}}.

\bibitem{CMS:2016ngn}
{CMS Collaboration}, {\em {The CMS trigger system}\/},
  \href{http://dx.doi.org/10.1088/1748-0221/12/01/P01020}{JINST {\bf 12} (2017)
  no.~01, P01020}, \href{http://arxiv.org/abs/1609.02366}{{\tt arXiv:1609.02366
  [physics.ins-det]}}.

\bibitem{LHCbCollaboration:2014vzo}
{LHCb Collaboration}, {\em {LHCb Trigger and Online Upgrade Technical Design
  Report}\/},  {LHCB-TDR-016}, 2014.
\newblock \url{http://cds.cern.ch/record/1701361}.

\bibitem{Colombo:2018upq}
T.~Colombo et al., {\em {FLIT-level InfiniBand network simulations of the DAQ
  system of the LHCb experiment for Run-3}\/},
  \href{http://dx.doi.org/10.1109/TNS.2019.2905993}{IEEE Trans. Nucl. Sci. {\bf
  66} (2019) no.~7, 1159--1164}, \href{http://arxiv.org/abs/1806.09527}{{\tt
  arXiv:1806.09527 [cs.NI]}}.

\bibitem{Behnke:2013lya}
H.~Abramowicz et al., {\em {The International Linear Collider Technical Design
  Report - Volume 4: Detectors}\/},  \href{http://arxiv.org/abs/1306.6329}{{\tt
  arXiv:1306.6329 [physics.ins-det]}}.

\bibitem{Collaboration:2759072}
{CMS Collaboration}, {\em {The Phase-2 Upgrade of the CMS Data Acquisition and
  High Level Trigger}\/},  {CMS-TDR-022}, 2021.
\newblock \url{https://cds.cern.ch/record/2759072}.

\end{thebibliography}\endgroup

\end{document}